\documentclass[reprint, amsmath, amssymb, aps, pra, floatfix]{revtex4-2}

\usepackage{subfigure}
\usepackage{amsmath}
\usepackage{enumitem}
\usepackage{graphicx}
\usepackage{dcolumn}
\usepackage{bm}
\usepackage[colorlinks=true,allcolors=blue]{hyperref}
\usepackage{booktabs} 
\usepackage{makecell}

\begin{document}

\title{Enhanced charging in open Rabi and Dicke quantum batteries}

\author{A-Long Zhou}

\author{Nuo Xu, Ya-Wen Xiao, Li-Li Gao}

\author{Long-Jie Li}

\author{Zi-Min Li}
\email{zimin.li@csu.edu.cn}

\author{Chuan-Cun Shu}
\email{cc.shu@csu.edu.cn}

\affiliation{Institute of Quantum Physics, Hunan Key Laboratory of Nanophotonics and Devices, Hunan Key Laboratory of Super-Microstructure and Ultrafast Process, School of Physics, Central South University, Changsha 410083, China}

\date{September 15, 2026}

\begin{abstract}
We investigate open-system charging protocols for Rabi and Dicke quantum batteries under direct incoherent pumping.
The charging dynamics follows a Lindblad master equation, while effective non-Hermitian generators describe the conditioned no-jump mechanism.
For a single Rabi battery, this pumping suppresses coherent energy backflow, broadens the high-energy charging window, and improves robustness against both timing errors and detuning.
At the selected Rabi operating points, pumping also increases ergotropy and the extractable fraction of the stored energy.
For finite Dicke batteries, direct incoherent pumping likewise stabilizes the charging dynamics.
Across the finite-size scan, the strongest pump considered improves the peak stored energy and ergotropy per cell and near-peak timing robustness relative to the Hermitian reference.
The maximum average charging power also increases, although this metric includes pump-assisted jump contributions.
Equal-input comparisons show that the relative performance of common-cavity Dicke and independent Rabi batteries varies with the pump-input bound.
Taken together, these results show that direct incoherent pumping provides a route to stabilized charging in Rabi and finite Dicke quantum batteries.
\end{abstract}

\maketitle

\section{Introduction}
\label{sec:intro}
Quantum batteries (QBs) have attracted broad interest as a setting in which quantum coherence and many-body effects may influence energy storage and transfer \cite{alicki_entanglement_2013, Felix_Quantacell_2015, Campaioli_Enhancing_2018, RevModPhys.96.031001, li_switchable_2025, shi_entanglement_2022}.
Cavity quantum electrodynamics provides a transparent architecture in which a cavity mode charges either a single two-level system (TLS) or an ensemble of TLSs \cite{andolina_quantum_2019, dou_extended_2022, seidov_quantum_2024, PhysRevA.103.023719, PhysRevA.103.013711, PhysRevA.104.033712, Li_2015, Xie_2017, Li_2021}.
The quantum Rabi model is the minimal setting in which a TLS battery exchanges energy coherently with a single bosonic charger mode, making it a useful benchmark for resolving the basic charging dynamics \cite{Le_Spin_2018, James_Superabsorption_2022}.
Its Dicke extension couples multiple TLS batteries to the same bosonic mode and thereby provides a controlled many-body benchmark for collective light--matter charging \cite{Ferraro_High_2018, Andolina_Extractable_2019, Juli_Bounds_2020, Caravelli_Random_2020, crescente_ultrafast_2020, PhysRevLett.134.130401, mayo_collective_2022, fusco_work_2016, Lu_2021}.
Here, the Rabi battery denotes a cavity system containing one TLS, whereas the Dicke battery denotes a common-cavity system containing $N$ TLSs; the term ``many-body'' is reserved for collective or finite-size features of the latter.

At the same time, purely Hermitian charging protocols remain constrained by coherent energy exchange.
For closed light--matter batteries, recurrent oscillations induce energy backflow, so that the highest charging energies are typically transient and require accurate timing of the decoupling operation \cite{Gyhm_Quantum_2022, Kamin_Non_2020, Pirmoradian_Aging_2019, Quach_Using_2020}.
The same coherent exchange also makes the performance sensitive to frequency detuning \cite{crescente_ultrafast_2020, Zhang_Powerful_2019}.
In stronger coupling regimes, counter-rotating processes further complicate the dynamics and can reduce the extractable fraction of the stored energy \cite{Ferraro_High_2018, PhysRevB.98.205423}.
For the Dicke battery, collective power enhancement can coexist with a suppression of the stored energy per cell, making the overall charging advantage less straightforward than the power scaling alone may suggest \cite{Ferraro_High_2018, Andolina_Extractable_2019}.
These features motivate the search for charging mechanisms that go beyond strictly unitary protocols.

Open-system driving provides one such route.
Direct incoherent pumping can continuously bias the battery toward higher-energy sectors and thereby reshape the balance between forward charging and coherent backflow.
Previous non-Hermitian quantum-battery studies have examined charging Hamiltonians with parity-time ($\mathcal{PT}$) or rotation-time symmetry, reservoir-engineered nonreciprocal transfer, $\mathcal{PT}$-symmetric wireless energy extraction, and topology-dependent charging in a Su--Schrieffer--Heeger (SSH) lattice \cite{PhysRevA.109.042207, lin2026enhanced, 8xsm-5mb6, 4klp-kw27}.
Here, we study Rabi and Dicke cavity batteries in which direct incoherent pumping acts on the TLS degrees of freedom and the cavity mediates coherent energy exchange.
The Lindblad master equation describes the physical charging dynamics, while the associated conditioned no-jump generators supply auxiliary descriptions of the pump-induced bias that are benchmarked only for selected battery observables \cite{hovhannisyan_charging_2020, Ashida02072020, RoccatiNon-Hermitian2022, PhysRevA.110.023708}.

Our main result is that direct incoherent pumping can stabilize the high-energy charging response that is otherwise limited by recurrent cavity--battery exchange.
For the Rabi battery, pumping suppresses coherent energy backflow, broadens the high-energy charging window, and reduces sensitivity to the stopping time and detuning.
We benchmark the conditioned no-jump dynamics against the Lindblad evolution and identify the observable-dependent region in which it provides a useful auxiliary description.
In a separate Rabi operating-point comparison, we use the ergotropy to examine the extractable component of the stored energy under an explicit input-bookkeeping convention.

For the Dicke battery, we combine representative $N=10$ dynamics with a finite-size scan to determine how the response to local incoherent pumping changes with battery size.
The finite-size scan shows observable-dependent tradeoffs at the two lower tested pump strengths, whereas at the largest tested value the stored energy, ergotropy, and timing robustness remain above the Hermitian values throughout the tested size range.
We then compare the common-cavity Dicke battery with $N$ independent Rabi batteries and find that their ordering depends on the accumulated pump-input bound.

The remainder of this paper is organized as follows.
We introduce the open-system protocol and performance metrics in Sec.~\ref{sec:model}.
Section~\ref{sec:rabi-results} analyzes the physical Lindblad charging dynamics of the Rabi battery and then separates the associated conditioned mechanism into a dedicated subsection.
Section~\ref{sec:thermo} examines extractable work in a separate Rabi operating-point comparison.
Section~\ref{sec:dicke-results} presents the representative Dicke dynamics, their finite-size dependence, the associated conditioned mechanism, and the equal-input comparison with independent Rabi batteries.
Section~\ref{sec:conclusion} summarizes the main results.
The detailed no-jump derivations, statistical diagnostics, and analytical reductions are collected in the appendices.

\section{Open-system protocol and performance metrics}
\label{sec:model}

\subsection{Physical charging protocol}

\begin{figure}[t]
    \centering
    \includegraphics[width=\linewidth]{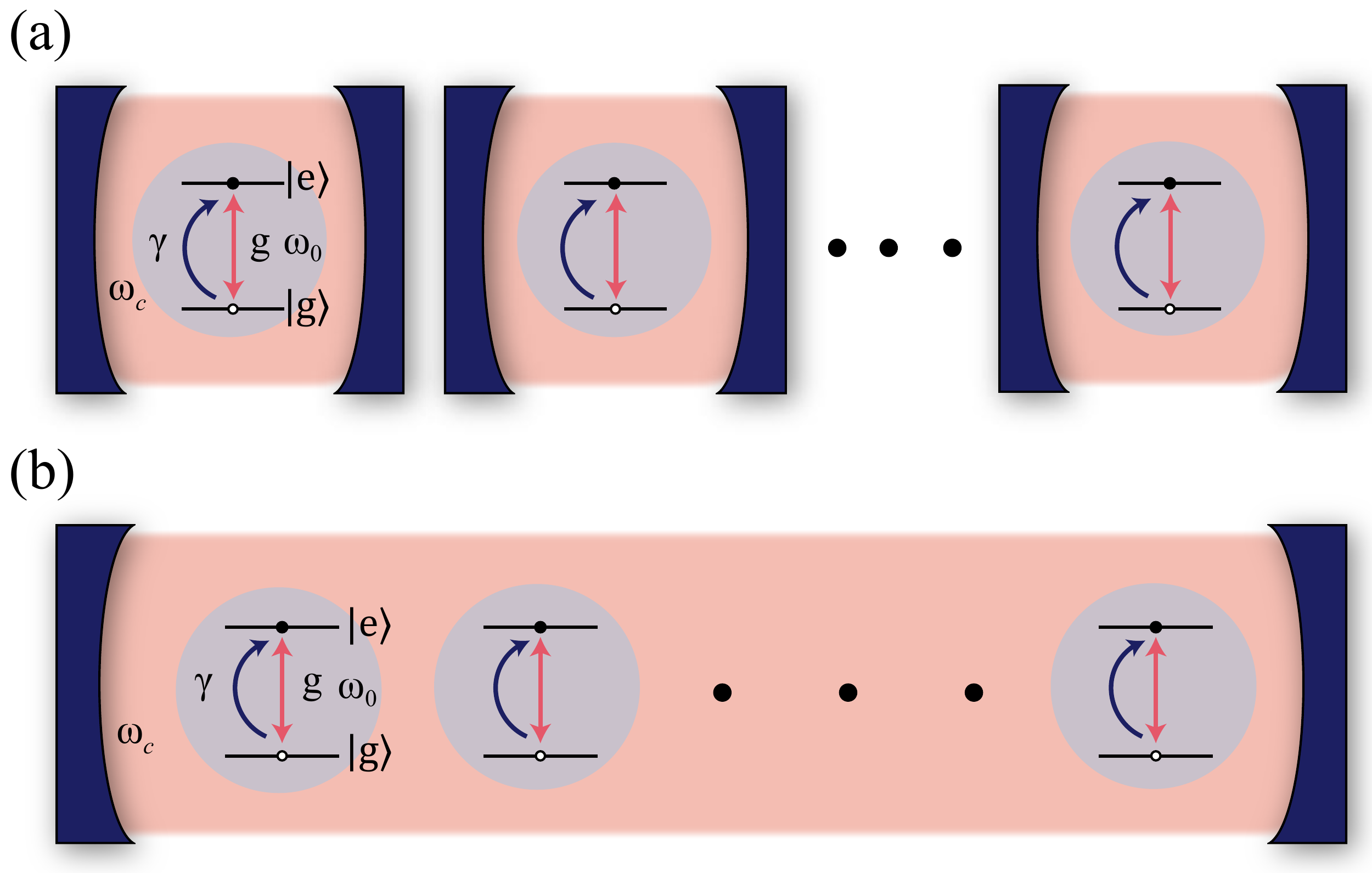}
    \caption{Schematic illustration of the open quantum battery models under direct incoherent pumping.
    (a) Independent Rabi reference composed of identical units, each containing one cavity mode and one TLS.
    The units have separate cavities and local pumps and do not interact.
    Each unit has cavity frequency $\omega_c$, TLS transition frequency $\omega_0$, light--matter coupling strength $g$, and local pump parameter $\gamma$.
    (b) Dicke battery in which all TLSs couple to one common cavity mode and receive identical local pumps.
    The Hamiltonian contains no direct interaction among the TLSs; the common cavity mediates their correlations.
    In both panels, the curved arrows denote the local incoherent excitation channels.}
    \label{fig:open-battery-schematic}
\end{figure}

We consider the cavity-battery configurations illustrated in Fig.~\ref{fig:open-battery-schematic}.
The physical protocol studied throughout this paper is the Lindblad evolution of these systems under direct incoherent pumping of the battery degree of freedom.
Within this framework, we analyze the Rabi battery and its Dicke extension.
We set $\hbar=1$ in the dynamical equations and restore it in the reported energy and power units.
For the Rabi battery, this process excites the TLS through $L=\sqrt{2\gamma}\sigma_+$, where $\gamma$ is the local pump parameter.
Here $\sigma_+=|e\rangle\langle g|$ and $\sigma_-=|g\rangle\langle e|$ are the TLS raising and lowering operators, respectively, with $|g\rangle$ and $|e\rangle$ denoting the ground and excited states.
For the Dicke battery, the same local excitation process acts separately on each TLS through $L_k=\sqrt{2\gamma}\sigma_+^{(k)}$, $k=1,\ldots,N$.
For either model, the pump contribution to the battery-energy current is $J_{\rm pump}^{(B)}(t)=2\gamma\omega_0\sum_k p_{g,k}(t)\geq0$, where $\omega_0$ is the TLS transition frequency and $p_{g,k}(t)$ is the ground-state population of TLS $k$; for the Rabi battery, the sum contains one term.
When the cavity contribution becomes negative and returns energy from the battery, this positive pump current reduces the net decrease of the stored energy; this reduced discharge is what we call suppression of backflow.
As $\gamma$ increases, the same current balance produces the broad energy plateau.
Appendix~\ref{app:dicke-energy-flow} relates the cavity current to the joint cavity--TLS exchange correlation and derives the explicit damping induced by the local pumps.

The operators $a$ and $a^\dagger$ annihilate and create a photon in the cavity mode, respectively.
For the Rabi battery, $H_{\rm R}=\omega_c a^\dagger a+\omega_0\sigma_z/2+g(a^\dagger+a)\sigma_x$, where $\omega_c$ and $\omega_0$ are the cavity and TLS frequencies, $g$ is the light--matter coupling strength, and $\sigma_x=\sigma_++\sigma_-$.
We denote the cavity--TLS detuning by $\Delta=\omega_0-\omega_c$.
At the density-matrix level, the Rabi battery obeys the Lindblad master equation
\begin{align}
    \frac{d\rho}{dt}
    =
    -i[H_{\rm R},\rho]
    +\mathcal D[L]\rho,
\end{align}
where $\mathcal D[L]\rho=L\rho L^\dagger-\frac{1}{2}\{L^\dagger L,\rho\}$.
Throughout this work, the total Hilbert space is ordered as $\mathcal H_c\otimes\mathcal H_B$, with the cavity factor written first.
The Rabi basis states are $|n\rangle_c\otimes|g\rangle_B$ and $|n\rangle_c\otimes|e\rangle_B$.
We write Dicke states as $|n\rangle_c\otimes|j,m\rangle_B$, where $j$ and $m$ are the total-spin and spin-projection quantum numbers; the fully symmetric initial state used below has $j=N/2$.
The initial density matrix $\rho(0)$ is fixed by the charging protocol.
Unless stated otherwise, the Rabi battery starts from $|1\rangle_c\otimes|g\rangle_B$, while the Dicke battery starts from $|N\rangle_c\otimes|j,-j\rangle_B$.
In the operating-point comparison of Sec.~\ref{sec:thermo}, the Hermitian reference starts from $|1\rangle_c\otimes|g\rangle_B$, whereas the pumped protocols start from $|0\rangle_c\otimes|g\rangle_B$.
For the Dicke battery, the collective spin operators are $J_\alpha=\frac{1}{2}\sum_{k=1}^{N}\sigma_\alpha^{(k)}$, with $\alpha=x,y,z$, and its Lindblad master equation is
\begin{align}
    \frac{d\rho}{dt}
    =
    -i[H_{\rm D},\rho]
    +
    2\gamma\sum_{k=1}^{N}\mathcal D[\sigma_+^{(k)}]\rho ,
\end{align}
and the Dicke Hamiltonian is
\begin{align}
    H_{\rm D}
    =
    \omega_c a^\dagger a+\omega_0 J_z+2g(a^\dagger+a)J_x .
\end{align}
This Lindblad dynamics with local incoherent pumping is permutation invariant.
The same jump operators also define conditioned no-jump representations, but these auxiliary descriptions are not needed to specify the physical protocol or its observables.
Their mechanism-level roles are examined separately for the Rabi and Dicke batteries in Secs.~\ref{sec:nh-mechanism} and \ref{sec:dicke-nh-mechanism}, while Appendix~\ref{DENHH} gives the detailed derivations.

Figure~\ref{fig:open-battery-schematic}(a) represents $N$ independent copies of the Rabi battery, each with its own cavity mode and local pump.
The units have no interaction and remain uncorrelated for the product initial states considered here.
Their dynamics therefore consists of $N$ copies of the Rabi calculation, and this arrangement provides the independent reference used in Sec.~\ref{sec:equal-input-comparison}.

\subsection{Battery observables and performance metrics}
For the physical Lindblad evolution, the reduced density matrix of the battery subsystem is
\begin{align}
    \rho_B(t) = \mathrm{Tr}_c\rho(t),
\end{align}
where $\text{Tr}_c$ denotes the partial trace over the cavity degrees of freedom.

The charging energy is defined as the energy increase of the battery subsystem,
\begin{align}
    \Delta E(t) = \text{Tr}[H_B \rho_B(t)] - \text{Tr}[H_B \rho_B(0)],
\end{align}
where $H_B = \frac{\omega_0}{2}\sigma_z$ for the Rabi model and $H_B = \omega_0 J_z$ for the Dicke model.
For $t>0$, the average charging power is defined as
\begin{align}
    \Delta P(t) = \frac{\Delta E(t)}{t}.
\end{align}
When the right-hand limit is finite, we use the continuous extension $\Delta P(0)\equiv\lim_{t\to0^+}\Delta P(t)$.

To characterize the optimal charging performance, we further introduce the maximum charging energy and the maximum average charging power.
The maximum charging energy is defined as
\begin{align}
\Delta E_{\max}=\max_t \Delta E(t),
\end{align}
and the maximum average charging power as
\begin{align}
\Delta P_{\max}=\max_{t\geq0}\Delta P(t).
\end{align}
Here, \(\Delta E_{\max}\) characterizes the largest energy stored during the charging process, whereas \(\Delta P_{\max}\) measures the highest average charging rate.
For the Lindblad protocol, \(\Delta P(t)\) is the average rate of battery-energy gain generated jointly by the cavity coupling and pump jumps.
Appendix~\ref{app:dicke-energy-flow} separates these two contributions.

\begin{figure*}[t]
    \centering
    \includegraphics[width=\linewidth]{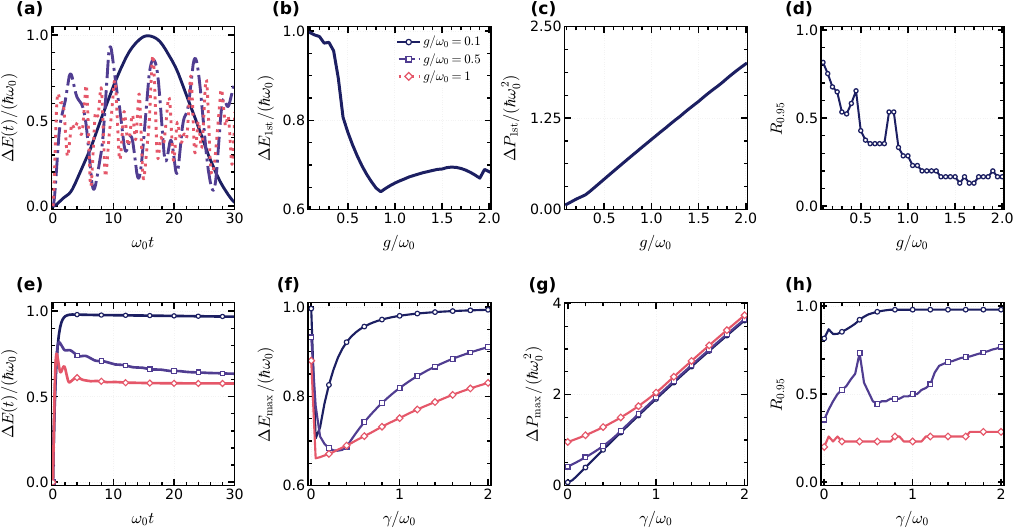}
    \caption{Rabi battery charging benchmarks for the Hermitian reference and full-Lindblad protocol at exact resonance and fixed initial state $|\psi(0)\rangle=|1\rangle_c\otimes|g\rangle_B$.
    Panels (a)--(d) show the Hermitian reference: (a) charging dynamics for $g/\omega_0=0.1,0.5,1$, (b) first-cycle charging energy, (c) first-cycle average power, and (d) timing robustness $R_{0.95}$ as functions of $g/\omega_0$.
    Panels (e)--(h) show the physical Lindblad protocol: (e) charging dynamics at fixed $\gamma/\omega_0=1$ for the same three couplings, followed by (f) maximum stored energy, (g) maximum average charging power, and (h) $R_{0.95}$ as functions of $\gamma/\omega_0$.
    In the discrete three-coupling comparisons, colors encode $g/\omega_0=0.1,0.5,1$ from purple to red; line styles in panel (a) and circle, square, and diamond markers in the lower row provide redundant identification.
    Panels (f)--(h) connect 41 calculated values $\gamma/\omega_0=0,0.05,\ldots,2$, with markers shown at intervals.
    The first-cycle notation in panels (b) and (c) refers to the first charging maximum of the Hermitian reference, whereas panels (f) and (g) use the peak quantities defined in Sec.~\ref{sec:model}.
    The Lindblad maximum average charging power includes both cavity-mediated and pump-assisted energy input.
    Numerical cutoffs and sampling details are summarized in Appendix~\ref{app:numerical-controls}.}
    \label{fig:single-battery-benchmarks}
\end{figure*}

For the upper row representing the Hermitian reference in Fig.~\ref{fig:single-battery-benchmarks}, $W_{0.95}$ is the width of the contiguous interval containing the first-cycle maximum over which $\Delta E(t)$ remains above $95\%$ of that maximum.
For the Lindblad Rabi and Dicke curves, $W_{0.95}$ is the longest contiguous interval over which $\Delta E(t)$ remains above $95\%$ of its global maximum in the stated simulation window.
The normalized robustness is then
\begin{align}
R_{0.95}=\frac{\omega_0W_{0.95}}{1+\omega_0W_{0.95}}.
\end{align}
The width $W_{0.95}$ has units of time and can become large when the stored energy remains above its curve-specific threshold over a broad plateau.
We first measure it in the natural time unit $\omega_0^{-1}$, giving $x=\omega_0W_{0.95}$, and then apply the monotonic map $x/(1+x)$.
This map preserves the ordering of the raw window widths while placing the timing robustness in $[0,1)$: $R_{0.95}=0$ for a vanishing window and approaches unity as the near-peak window broadens.
Because the threshold is defined relative to each curve's own peak, $R_{0.95}$ measures timing tolerance near that peak and must be interpreted together with $\Delta E_{\max}$ rather than as robustness above a common absolute energy target.

Because the open dynamics can generate mixed battery states, the total stored energy and the extractable part of that energy need not coincide.
We therefore complement $\Delta E(t)$ by the ergotropy $\mathcal{W}(t)$, which measures the useful energy that can be converted into work by cyclic unitary operations \cite{alicki_entanglement_2013, allahverdyan_maximal_2004, Yang_Battery_2023}.
It is defined as
\begin{align}
    \mathcal{W}(t) = \text{Tr}[H_B \rho_B(t)] - \text{Tr}[H_B \rho_B^{\text{pass}}(t)].
\end{align}
Here $\rho_B^{\text{pass}}(t)$ is the passive state associated with $\rho_B(t)$.
This state is constructed by rearranging the eigenvalues of $\rho_B(t)$ in descending order ($r_0 \ge r_1 \ge \dots$) and assigning them to the eigenstates of $H_B$ in ascending energy order ($\epsilon_0 \le \epsilon_1 \le \dots$),
\begin{align}
\rho_B^{\mathrm{pass}}(t) = \sum_n r_n |\epsilon_n\rangle \langle \epsilon_n|.
\end{align}
For the single-TLS calculations, we evaluate $\mathcal W(t)$ by diagonalizing $\rho_B(t)$ and constructing the passive state as defined above.
In the basis $(|e\rangle_B,|g\rangle_B)$, we write
\begin{align}
\rho_B(t)
&=
\begin{pmatrix}
\rho_{ee}(t) & \rho_{eg}(t)\\
\rho_{ge}(t) & \rho_{gg}(t)
\end{pmatrix},
&
z(t)&=\rho_{ee}(t)-\rho_{gg}(t).
\end{align}
For $H_B=\omega_0\sigma_z/2$, the resulting ergotropy is
\begin{align}
\mathcal W(t)
=
\frac{\omega_0}{2}
\left[
z(t)+\sqrt{z^2(t)+4|\rho_{eg}(t)|^2}
\right].
\end{align}
When the reduced state is diagonal in the energy basis, this expression becomes $\mathcal W(t)=\omega_0\max[0,2\rho_{ee}(t)-1]$.
For the Dicke calculations, we use the general passive-state construction with the degeneracies of $H_B=\omega_0J_z$ included.

\section{Rabi battery charging under direct incoherent pumping}
\label{sec:rabi-results}
We first analyze the unpumped unitary Rabi dynamics generated by $H_{\rm R}$, which serves as the Hermitian reference, and identify its tradeoff among stored energy, average charging power, and timing robustness.
We then show how direct incoherent pumping reshapes the physical Lindblad charging dynamics over both the reference and stronger-coupling regimes.
Finally, we isolate the conditioned no-jump description and identify the parameter regions in which it remains a useful observable-level guide for the selected battery quantities.

\subsection{Hermitian reference and Lindblad stabilization}

Figure~\ref{fig:single-battery-benchmarks}(a) shows that the $g/\omega_0=0.1$ trace reaches near-complete charging, whereas larger $g$ produces increasingly structured dynamics and incomplete transfer.
The scan in Fig.~\ref{fig:single-battery-benchmarks}(b) and Fig.~\ref{fig:single-battery-benchmarks}(c) shows the corresponding tradeoff: the first-cycle charging energy decreases with $g$, while the first-cycle average power of the Hermitian reference increases.
The Hermitian reference therefore exhibits a tradeoff between stored energy and average charging power for the Rabi battery.
For the Hermitian reference, the first charging maximum is the natural operational stopping point of the unitary cycle.
Accordingly, the Hermitian reference is summarized by first-cycle quantities, whereas the open-system row uses the peak quantities defined in Sec.~\ref{sec:model}.

The same tradeoff appears in the timing-robustness diagnostic shown in Fig.~\ref{fig:single-battery-benchmarks}(d).
As $g$ increases, the high-energy window rapidly narrows and the normalized robustness $R_{0.95}$ decreases, showing that the faster Hermitian charging cycles are also more sensitive to timing errors.
The recurrent backflow visible in Fig.~\ref{fig:single-battery-benchmarks}(a) concentrates the stored energy near sharp transient maxima and prevents a broad plateau in the Rabi battery.
Because $R_{0.95}$ is evaluated from sampled threshold windows, Fig.~\ref{fig:single-battery-benchmarks}(d) shows calculated points joined only as a guide to the eye.

Since $g/\omega_0=0.1$ gives the largest first-cycle stored energy in the scan of the Hermitian reference, we use this reference coupling as the baseline for assessing how pumping changes the Rabi battery dynamics.
The lower row of Fig.~\ref{fig:single-battery-benchmarks} extends this comparison to the physical Lindblad protocol at $g/\omega_0=0.1$, $0.5$, and $1$ over 41 explicitly calculated pump strengths.
At the reference coupling, comparison of panels (a) and (e), together with the dense scans in panels (f)--(h), shows that pumping removes the pronounced recurrent discharge, produces a broad high-energy plateau, and increases the maximum average charging power.
Because actual pump jumps contribute to the Lindblad energy gain, panel (g) measures the total open-system charging rate rather than a purely cavity-mediated transfer rate.

At $\gamma/\omega_0=1$, comparison of the $g/\omega_0=0.5$ and $1$ traces in panels (a) and (e) shows that the recurrent energy return is strongly reduced, although their peak stored energies in panel (f) remain below the corresponding unpumped maxima.
The timing robustness $R_{0.95}$ increases from $0.355$ to $0.500$ at $g/\omega_0=0.5$ and from $0.200$ to $0.231$ at $g/\omega_0=1$.
Pumping therefore continues to suppress backflow at the tested stronger couplings, while the improvement in timing robustness becomes more modest as $g$ increases.

\subsection{Detuning robustness under direct incoherent pumping}
\begin{figure}[t]
    \centering
    \includegraphics[width=\linewidth]{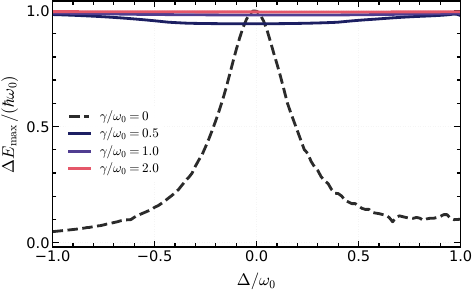}
    \caption{Maximum stored energy of the Rabi battery as a function of detuning at the reference coupling $g/\omega_0=0.1$.
    The dashed black curve is the Hermitian reference, and the colored solid curves are the physical Lindblad results for $\gamma/\omega_0=0.5,1,2$ over $\Delta/\omega_0\in[-1,1]$.
    Direct incoherent pumping replaces the sharp Hermitian resonance peak by a broad high-energy region.}
    \label{fig:detuning-robustness}
\end{figure}
To quantify the effect of direct incoherent pumping on detuning sensitivity, we compute the maximum charging energy $\Delta E_{\max}$ as a function of the detuning $\Delta$ at the reference coupling $g/\omega_0=0.1$.
As shown in Fig.~\ref{fig:detuning-robustness}, the Hermitian reference exhibits a narrow resonance peak, slightly shifted from exact resonance because the counter-rotating terms of the full Rabi model induce a Bloch-Siegert correction.
By contrast, the Lindblad curves remain close to a broad high-energy plateau across a much wider detuning window.
Pumping therefore substantially reduces the sensitivity of the Rabi battery to frequency mismatch.
The conditioned mechanism discussed next provides a complementary interpretation of this plateau without changing the Lindblad evidence on which this conclusion rests.

\subsection{Conditioned non-Hermitian mechanism}
\label{sec:nh-mechanism}

The Lindblad results above establish the physical charging response without invoking a non-Hermitian approximation.
To isolate the drift associated with records containing no pump jump, we use the effective Rabi generator
\begin{align}
H_{\rm R}^{\rm eff}
&=H_{\rm R}-\frac{i}{2}L^\dagger L \notag\\
&=H_{\rm R}+i\frac{\gamma}{2}\sigma_z-i\frac{\gamma}{2}\mathbb I .
\label{eq:rabi-effective-generator}
\end{align}
The state-independent term $-i\gamma\mathbb I/2$ changes the norm and hence the no-jump probability, but it cancels from instantaneously normalized observables.
The term $i\gamma\sigma_z/2$ favors the excited component relative to the ground component along the normalized no-jump record, thereby damping the conditional return of battery energy generated by coherent cavity--TLS exchange.
Appendix~\ref{DENHH} gives the full Rabi and Dicke derivations and shows explicitly which state-independent terms are removed.

Let $|\psi_{\rm nj}(t)\rangle$ denote the unnormalized state propagated with $H_{\rm R}^{\rm eff}$.
Battery observables along this record are evaluated from the instantaneously normalized reduced state
\begin{align}
\rho_B^{\rm nj}(t)
=
\operatorname{Tr}_c
\left[
\frac{|\psi_{\rm nj}(t)\rangle\langle\psi_{\rm nj}(t)|}
{\langle\psi_{\rm nj}(t)|\psi_{\rm nj}(t)\rangle}
\right].
\end{align}
This conditioned state is used only to interpret the mechanism; every performance conclusion in Figs.~\ref{fig:single-battery-benchmarks} and \ref{fig:detuning-robustness} is evaluated from the full Lindblad density matrix.

\begin{figure}[!t]
    \centering
    \includegraphics[width=0.92\linewidth]{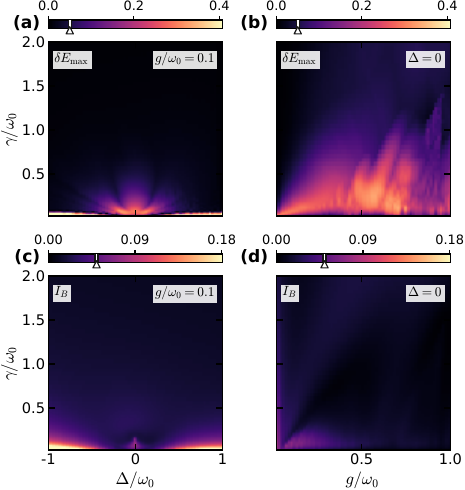}
    \caption{Observable-level usage map of the conditioned non-Hermitian description for the Rabi battery.
    Panels (a) and (b) show the peak-energy discrepancy $\delta E_{\max}=|\Delta E_{\max}^{\rm nj}-\Delta E_{\max}^{\rm ME}|$ in units of $\hbar\omega_0$ over the $(\Delta,\gamma)$ and $(g,\gamma)$ planes, respectively, where nj and ME denote the conditioned no-jump and Lindblad master-equation results.
    Panels (c) and (d) show the time-averaged reduced-battery-state infidelity $I_B$ between the same two evolutions over the corresponding parameter planes.
    The fidelity is sampled uniformly over $t_j\in[0,t_{\rm cut}]$, with $t_{\rm cut}\omega_0=50$ and $N_t=201$, after tracing out the cavity.
    In panels (a) and (c), $g/\omega_0=0.1$; in panels (b) and (d), $\Delta=0$.
    The two panels in each row share a color scale capped at the 98th percentile of their combined values; values above this limit are shown with the saturated endpoint color.
    White notches and open triangles mark $\delta E_{\max}=0.05\hbar\omega_0$ and $I_B=0.05$; lower-value colors guide the identification of low discrepancy in each observable rather than defining sharp boundaries.
    Appendix Fig.~\ref{fig:nojump-jump-sectors} separately reports the no-jump probability and compares the no-jump and jump-conditioned reduced battery states.}
    \label{fig:single-battery-applicability}
\end{figure}

Figure~\ref{fig:single-battery-applicability} identifies where this conditioned description reproduces selected Lindblad battery observables.
The peak-energy discrepancy $\delta E_{\max}=|\Delta E_{\max}^{\rm nj}-\Delta E_{\max}^{\rm ME}|$ tests agreement at the charging maximum, while the time-averaged reduced-battery-state infidelity
\begin{align}
I_B
&=
1-\frac{1}{N_t}\sum_j F_B(t_j),
&
F_B(t_j)
&=
F\!\left[\rho_B^{\rm nj}(t_j),\rho_B^{\rm ME}(t_j)\right]
\end{align}
tests reduced-state agreement over the simulation window.
Here $\rho_B^{\rm nj}$ and $\rho_B^{\rm ME}$ denote the conditioned no-jump and Lindblad battery states obtained after tracing out the cavity.
We use the Uhlmann root-fidelity convention $F(\rho,\sigma)=\operatorname{Tr}\sqrt{\sqrt{\rho}\sigma\sqrt{\rho}}$, consistent with the numerical evaluation.
In the detuning maps, the low-discrepancy region broadens rapidly as the pumping increases, whereas the coupling maps retain nonmonotonic higher-discrepancy structures at larger $g$.
The peak-energy and time-averaged state comparisons also need not select identical regions, so the conditioned description must be assessed for the observable of interest.
The markers at $\delta E_{\max}=0.05\hbar\omega_0$ and $I_B=0.05$ guide the identification of low discrepancy in each observable and do not define universal validity boundaries.

The analytical detuning interpretation in Appendices~\ref{ASEPD} and \ref{ENHH} further reduces the conditioned Rabi model to a non-Hermitian Jaynes--Cummings model.
This step uses the rotating-wave approximation, which near resonance requires $g\ll\omega_0,\omega_c$, or approximately $g/\omega_0\ll1$.
With $\widetilde{\Omega}_1=\sqrt{4g^2+(\Delta+i\gamma)^2}$, the late-time expression additionally requires $\operatorname{Im}\widetilde{\Omega}_1>0$, which becomes $\gamma>2g$ at exact resonance.
It therefore provides mechanism-level guidance at the reference coupling $g/\omega_0=0.1$ and is not used to extrapolate the quantitative strong-coupling results in Fig.~\ref{fig:single-battery-benchmarks}(e)--(h).

Agreement between normalized no-jump and Lindblad observables does not imply that the no-jump record has a large statistical weight.
Before the state-independent imaginary shift is removed, the squared norm gives the no-jump probability $P_{\rm nj}(t)$, and the reduced Lindblad battery state obeys the exact decomposition
\begin{align}
\rho_B^{\rm ME}(t)
=
P_{\rm nj}(t)\rho_B^{\rm nj}(t)
+
[1-P_{\rm nj}(t)]\rho_B^{\rm J}(t).
\end{align}
Here $\rho_B^{\rm nj}$ and $\rho_B^{\rm J}$ are the reduced battery states conditioned on no pump jump and on at least one pump jump, respectively.
At $\gamma/\omega_0=1$, $P_{\rm nj}=0.0125$ at the normalized no-jump energy maximum, whereas the fidelity between the no-jump and jump-conditioned reduced battery states is $0.9985$.
The two conditional sectors can therefore produce similar battery observables at this pump strength because both approach highly excited reduced battery states, even though the no-jump sector is rare.
This reduced-state agreement neither makes no-jump histories dominant in the Lindblad ensemble nor establishes equivalence of the full cavity--battery states.

\section{Extractable work in the Rabi battery}
\label{sec:thermo}
\begin{figure}[tb]
    \centering
    \includegraphics[width=0.9\linewidth]{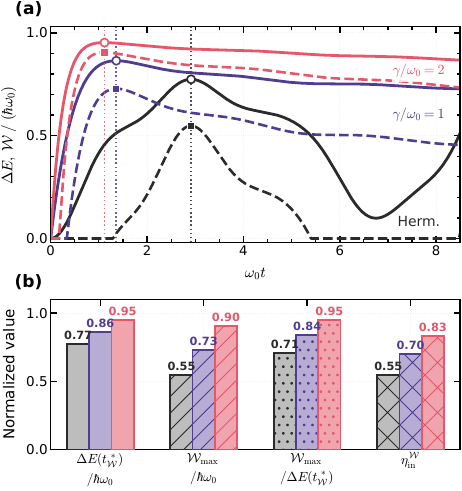}
    \caption{Operational comparison via ergotropy with input bookkeeping for the Rabi battery at $g/\omega_0=0.5$.
    In panel (a), solid and dashed curves denote the stored energy $\Delta E(t)$ and ergotropy $\mathcal{W}(t)$, respectively; black, purple, and red identify the Hermitian reference and the pumped protocols at $\gamma/\omega_0=1$ and $2$.
    Open circles and filled squares mark $\Delta E$ and $\mathcal{W}$ at the ergotropy-optimal time $t_{\mathcal{W}}^*$, also indicated by the vertical dotted lines.
    Panel (b) groups $\Delta E(t_{\mathcal{W}}^*)$, $\mathcal{W}_{\max}$, $\mathcal{W}_{\max}/\Delta E(t_{\mathcal{W}}^*)$, and the adopted input-normalized ergotropy yield $\eta_{\rm in}^{\mathcal W}$ by horizontal position; within each group the black--purple--red order identifies the protocols, and hatching provides a redundant metric cue.
    The first two quantities are normalized by $\hbar\omega_0$, whereas the latter two are dimensionless.
    The comparison uses the ergotropy-optimal operating point and does not impose a fixed input budget.}
    \label{fig:ergotropy-reference-cost}
\end{figure}

We compare the ergotropy $\mathcal{W}(t)$ defined in Sec.~\ref{sec:model} for the Hermitian reference, initialized with one photon in the cavity, and for pumped protocols initialized with an empty cavity.
For the comparison shown in Fig.~\ref{fig:ergotropy-reference-cost}, we use $g/\omega_0=0.5$, for which the separation between stored energy and extractable work is more clearly resolved within the plotted time window.
For each protocol, $t_{\mathcal W}^{*}$ denotes the time at which $\mathcal W(t)$ reaches its maximum within the time window considered here.

We adopt an input-bookkeeping convention for the protocols shown in Fig.~\ref{fig:ergotropy-reference-cost}.
The Hermitian reference starts from $|1\rangle_c\otimes|g\rangle_B$ and is powered by the one photon preloaded in the cavity.
The pumped protocols start from $|0\rangle_c\otimes|g\rangle_B$, so their input is supplied solely through the pump channel.
Adding the pump to the one-photon preparation instead defines a hybrid protocol that combines both resources.
For the pumped cases, we track the cumulative input from the pumping channel with respect to the Hermitian Rabi Hamiltonian $H_{\rm R}$,
\begin{align}
    Q_{\mathrm{in}}^{\mathrm{sys}}(t)
    =
    \int_0^t
    \mathrm{Tr}
    \left[
        H_{\rm R} \mathcal{D}_{\mathrm{pump}}[\rho(t')]
    \right]
    dt' .
\end{align}
Here $\mathcal D_{\rm pump}[\rho]$ is the pump contribution to $\dot\rho$ in this bookkeeping convention.
All pumped quantities below are evaluated at the ergotropy-optimal operating time $t_{\mathcal{W}}^*$.
For the pumped curves in Fig.~\ref{fig:ergotropy-reference-cost}, the accumulated inputs at $t_{\mathcal{W}}^*$ are $Q_{\mathrm{in}}^{\mathrm{sys}}(t_{\mathcal{W}}^*)/\hbar\omega_0=1.043$ and $1.084$ for $\gamma/\omega_0=1$ and $2$, respectively.
The comparison does not impose a fixed input budget or choose the stopping time from such a budget.
It normalizes the output at the selected ergotropy-optimal time by the accumulated input in the adopted bookkeeping convention.
The resulting comparison is therefore an input-normalized diagnostic of the selected operating point and is not interpreted as a complete thermodynamic efficiency of the pump reservoir.

Figure~\ref{fig:ergotropy-reference-cost}(a) compares the time evolution of the stored energy and the ergotropy.
The Hermitian reference exhibits slow oscillatory dynamics, and a visible gap opens between $\Delta E(t)$ and $\mathcal{W}(t)$.
This separation shows that part of the stored energy remains locked in the passive component of the battery state.

The incoherent pumping protocols accelerate charging and, at their respective ergotropy-optimal times $t_{\mathcal W}^{*}$, retain a larger extractable fraction than the Hermitian reference.
At $\gamma/\omega_0=2$, the ergotropy nearly reaches the stored energy at the marked ergotropy peak time $t_{\mathcal{W}}^*$.
For the Rabi operating points considered here, pumping therefore increases both the stored energy and the extractable fraction retained in the battery relative to the stated Hermitian reference.

Figure~\ref{fig:ergotropy-reference-cost}(b) summarizes the same ergotropy-optimal operating points through four quantities.
The first three groups show the stored energy $\Delta E(t_{\mathcal{W}}^*)$, the ergotropy $\mathcal{W}(t_{\mathcal{W}}^*)=\mathcal{W}_{\max}$, and the extractable-work fraction $\mathcal{W}(t_{\mathcal{W}}^*)/\Delta E(t_{\mathcal{W}}^*)$.
The fourth group shows the adopted input-bookkeeping yield
\begin{align}
    \eta_{\mathrm{in}}^{\mathcal{W}}
    =
    \frac{\mathcal{W}(t_{\mathcal{W}}^*)}
    {Q_{\mathrm{in}}^{\mathrm{sys}}(t_{\mathcal{W}}^*)}.
\end{align}
For the Hermitian reference, the denominator is the one-photon input supplied initially by the cavity.
Within this bookkeeping convention, the incoherent pumping cases yield larger ergotropy, a higher extractable-work fraction, and a larger value of this adopted input-bookkeeping yield than the Hermitian reference.
This operating-point comparison supports the same qualitative conclusion as Fig.~\ref{fig:ergotropy-reference-cost}(a): under the stated preparation and bookkeeping convention, pumping can turn a larger fraction of the stored energy into extractable work.

\section{Local incoherent pumping in the Dicke battery}
\label{sec:dicke-results}
We now turn to the Dicke battery.
We first examine how local incoherent pumping modifies representative $N=10$ dynamics relative to the Hermitian reference, then test the finite-size dependence of the resulting performance metrics, isolate the associated conditioned mechanism, and finally compare the common-cavity architecture with independent Rabi batteries under an equal accumulated pump-input bound.

\subsection{Response to local incoherent pumping at fixed size}

\begin{figure}[tb]
    \centering
    \includegraphics[width=0.94\columnwidth]{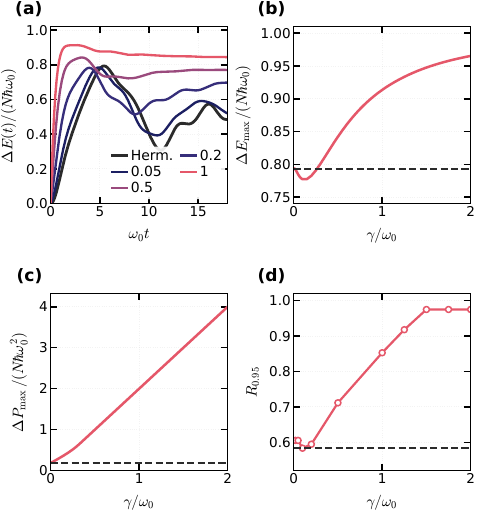}
    \caption{Response of a representative Dicke battery to local incoherent pumping at $N=10$ and $g/\omega_0=0.1$, initialized in $|N\rangle_c\otimes|j,-j\rangle_B$.
    Panel (a) shows the charging dynamics for the Hermitian reference and selected Lindblad protocols with local incoherent pumping; numerical legend entries denote $\gamma/\omega_0$.
    Panels (b)--(d) show the peak stored energy per cell, maximum average charging power per cell, and timing robustness over $0\le\gamma/\omega_0\le2$.
    Horizontal dashed lines mark the corresponding Hermitian values.
    The maximum average charging power contains cavity-mediated and pump-assisted contributions; the reported value includes the analytical $t\rightarrow0^+$ limit $\Delta P/(N\hbar\omega_0^2)=2\gamma/\omega_0$.
    Numerical details are given in Appendix~\ref{app:numerical-controls}.}
    \label{fig:dicke-direct-pumping-response}
\end{figure}

Figure~\ref{fig:dicke-direct-pumping-response} summarizes the representative Dicke benchmark at $N=10$ and $g/\omega_0=0.1$.
We compare the Hermitian reference with the Lindblad protocol using identical local pumps and the same battery observables as in the Rabi analysis: charging dynamics, peak stored energy, maximum average charging power, and timing robustness.
In both cases, the benchmark starts from the cavity-loaded state $|N\rangle_c\otimes|j,-j\rangle_B$, namely $N$ cavity photons with all TLSs in their ground state, so that the Hermitian reference and the pumped benchmark share the same common-cavity charging resource at the initial time.

Figure~\ref{fig:dicke-direct-pumping-response}(a) shows that the Hermitian reference for the Dicke battery undergoes recurrent charging and discharging, with substantial backflow after the first charging maximum.
Local incoherent pumping suppresses this backflow and turns the later-time dynamics into a smoother high-energy response.
At $\gamma/\omega_0=0.2$, the oscillations are mainly reshaped but remain visible, whereas the traces at $\gamma/\omega_0=0.5$ and $1$ approach a stable high-energy plateau.

In Fig.~\ref{fig:dicke-direct-pumping-response}(b), the peak stored energy initially changes little from its value at $\gamma=0$ and then rises above its value for the Hermitian reference as the pumping increases.
Figure~\ref{fig:dicke-direct-pumping-response}(c) shows a stronger increase in the maximum average charging power.
The energy-flow decomposition in Appendix~\ref{app:dicke-energy-flow} shows that the finite-time maximum at $\gamma/\omega_0=0.1$ contains comparable cavity and pump contributions, whereas the maxima at $\gamma/\omega_0=0.5$ and $1$ are set by the initial pump input.
The increased maximum average charging power therefore characterizes the total charging protocol, including direct pump input.
It does not by itself establish an intrinsic cavity-mediated or collective charging advantage.
Figure~\ref{fig:dicke-direct-pumping-response}(d) shows that the timing robustness $R_{0.95}$ improves at $\gamma/\omega_0=0.5$ and $1$, indicating that the high-energy state can be maintained over a wider stopping-time window.

\subsection{Finite-size dependence}

\begin{figure}[tb]
    \centering
    \includegraphics[width=0.94\columnwidth]{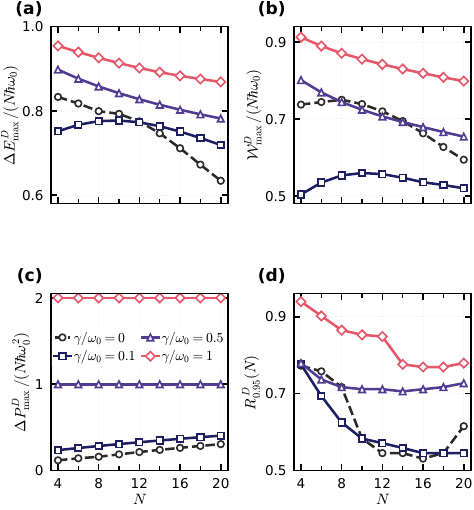}
    \caption{Finite-size response of the locally pumped Dicke battery at $g/\omega_0=0.1$, initialized in $|N\rangle_c\otimes|j,-j\rangle_B$.
    Panels (a)--(d) show the peak stored energy, peak ergotropy, maximum average charging power per cell, and timing robustness, respectively; the superscript $D$ labels the Dicke architecture.
    Black circles, dark-indigo squares, purple triangles, and red diamonds denote $\gamma/\omega_0=0,0.1,0.5,1$ for $N=4,6,\ldots,20$; connecting lines are guides to the eye.
    The ergotropy uses the degeneracy-aware passive-state construction.
    The maximum average charging power contains cavity-mediated and pump-assisted contributions and includes the analytical $t\rightarrow0^+$ limit.
    Cutoffs and sampling details are given in Appendix~\ref{app:numerical-controls}.}
    \label{fig:dicke-finite-size}
\end{figure}

To determine how the representative $N=10$ behavior changes with battery size, we perform the finite-size scan shown in Fig.~\ref{fig:dicke-finite-size}.

Panels (a) and (b) show that the peak stored energy and peak ergotropy respond differently to local pumping.
At $\gamma/\omega_0=0.1$, the peak stored energy lies below the unpumped result for $N\leq12$ and above it for $N\geq14$, whereas the peak ergotropy remains lower throughout the scan.
At $\gamma/\omega_0=0.5$, the peak stored energy remains above the unpumped result, while the peak ergotropy is higher at the smallest and largest sizes but slightly lower at intermediate $N$.
For $\gamma/\omega_0=1$, both quantities remain above their unpumped values over the full size range.

Panel (c) shows that local pumping increases the maximum average charging power at every displayed $N$.
This quantity includes energy supplied through pump-induced quantum jumps and therefore measures the total open-system charging rate.
Panel (d) shows no systematic improvement of the timing robustness at $\gamma/\omega_0=0.1$, whereas $\gamma/\omega_0=1$ maintains a broader high-energy window throughout the scan.

Taken together, these trends show that the effect of local pumping depends on both the pumping strength and the chosen performance measure; at $\gamma/\omega_0=1$, the peak stored energy, peak ergotropy, and timing robustness near each curve's own peak remain above those of the Hermitian reference throughout the finite-size scan.

The different trends in Figs.~\ref{fig:dicke-finite-size}(a) and (b) are consistent with stored energy and ergotropy characterizing different properties of the reduced battery state.
The individual pump jumps increase the excited-state populations, while averaging over unobserved jump events can also mix the battery state and distribute weight among different total-spin sectors.
Such mixing would leave a larger fraction of the battery energy in the passive component, which cannot be extracted through cyclic unitary operations.
The reduced extractable fraction at $\gamma/\omega_0=0.1$ and its recovery at $\gamma/\omega_0=1$ are therefore consistent with stronger mixing at the lower pump strength and a larger population inversion at the higher pump strength.

\subsection{Conditioned Dicke mechanism}
\label{sec:dicke-nh-mechanism}

The Lindblad dynamics in Figs.~\ref{fig:dicke-direct-pumping-response} and \ref{fig:dicke-finite-size} remains the physical benchmark for the Dicke battery.
Along a no-jump record of the same local pumps, the effective Dicke generator is
\begin{align}
H_{\rm D}^{\rm eff}
&=H_{\rm D}-\frac{i}{2}\sum_{k=1}^{N}L_k^\dagger L_k \notag\\
&=H_{\rm D}+i\gamma J_z-i\frac{\gamma N}{2}\mathbb I .
\label{eq:dicke-effective-generator}
\end{align}
Normalization removes the state-independent final term, leaving the relative bias $i\gamma J_z$, which favors collective states with larger battery excitation.
Within the single-excitation block of the rotating-wave reduction, coherent exchange through the common cavity has the reference collective scale $g\sqrt{N}$.
Together with the conditioned excitation bias, this low-excitation result gives an auxiliary picture of their competition.
Because the benchmark starts from $N$ cavity photons and jump-containing trajectories remain part of the Lindblad ensemble, this picture is not used to infer an $N$-scaling law for the Lindblad stored energy or charging power, or a resource-independent advantage between architectures.
The no-jump derivation and the technical Tavis--Cummings reduction are given in Appendix~\ref{DENHH}.

The comparison above concerns the locally pumped Dicke battery and its Hermitian reference at the same $N$ and initial cavity loading.
In Fig.~\ref{fig:dicke-finite-size}, the same pump parameter $\gamma$ is applied to every TLS, so the sum of the bare local dissipator coefficients scales as $2N\gamma$.
We next compare the common-cavity and independent arrangements under the same accumulated pump-input bound.

\subsection{Equal-input comparison}
\label{sec:equal-input-comparison}

\begin{figure}[tb]
    \centering
    \includegraphics[width=0.94\columnwidth]{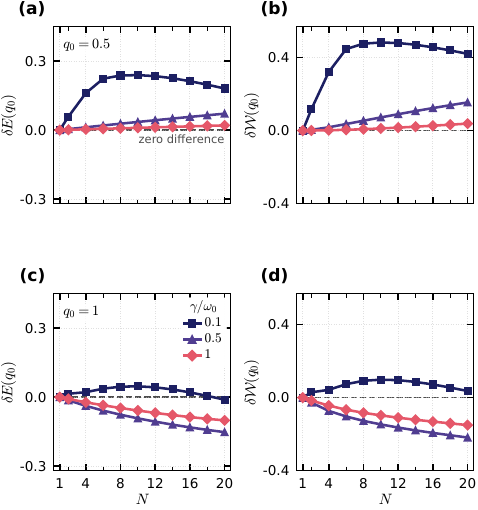}
    \caption{Comparison at the normalized accumulated pump-input bounds $q_0=0.5$ and $1$, with the first crossing of $q_X(t)=Q_{{\rm in},X}^{\rm sys}(t)/(N\hbar\omega_0)$ at $q_0$ defining the allowed time interval for each architecture.
    Panels (a) and (b) show the budget-constrained per-cell stored-energy and ergotropy differences at $q_0=0.5$, while panels (c) and (d) show the corresponding differences at $q_0=1$.
    Positive values favor the common-cavity Dicke battery, and the gray dashed horizontal lines mark zero difference between the two architectures.
    Results are shown for $N=1,2,4,6,\ldots,20$ and $\gamma/\omega_0=0.1,0.5,1$; dark-indigo squares, purple triangles, and red diamonds denote the three pump strengths.
    Each starred quantity is the maximum reached before the corresponding arrangement first attains the stated input bound.
    The differences vanish at $N=1$ because the two arrangements then describe the same Rabi battery.}
    \label{fig:dicke-independent-equal-input}
\end{figure}

We compare the common-cavity Dicke battery with $N$ independent Rabi batteries at the same total initial photon energy.
We use $D$ and $R$ to label the common-cavity Dicke and independent-Rabi arrangements, respectively.
The Dicke battery starts from $|N\rangle_c\otimes|j,-j\rangle_B$, whereas the independent arrangement starts from $\bigotimes_{k=1}^{N}\bigl(|1\rangle_{c_k}\otimes|g\rangle_{B_k}\bigr)$, so that
\begin{align}
E_{\rm cav}^{D}(0)=E_{\rm cav}^{R}(0)=N\hbar\omega_c .
\end{align}
For each arrangement $X=D,R$, we evaluate $Q_{{\rm in},X}^{\rm sys}(t)$ from the full-system pump-energy current defined in Sec.~\ref{sec:thermo}, using the corresponding full Hamiltonian and total local-pump dissipator, and define
\begin{align}
q_X(t)=\frac{Q_{{\rm in},X}^{\rm sys}(t)}{N\hbar\omega_0}.
\end{align}
The independent-Rabi input $Q_{{\rm in},R}^{\rm sys}$ sums the pump input over all $N$ cells.
We denote by $t_{q_0}^{X}$ the first time at which $q_X(t)$ reaches the input bound $q_0$.
All cases shown in Fig.~\ref{fig:dicke-independent-equal-input} reach both $q_0=0.5$ and $1$ within the simulation window.
The quantities $\Delta E_R$ and $\mathcal W_R$ below denote the corresponding single-cell observables, equivalently the additive array totals divided by $N$.

Because coherent backflow can lower an observable at the boundary of the allowed interval, we compare the largest stored energy and ergotropy reached before each arrangement first reaches the input bound,
\begin{align}
\Delta E_X^{\star}(q_0)&=\max_{0\leq t\leq t_{q_0}^X}\Delta E_X(t),\\
\mathcal W_X^{\star}(q_0)&=\max_{0\leq t\leq t_{q_0}^X}\mathcal W_X(t).
\end{align}
The per-cell differences shown below are
\begin{align}
\delta E(q_0)&=\frac{\Delta E_D^{\star}(q_0)/N-\Delta E_R^{\star}(q_0)}{\hbar\omega_0},\\
\delta \mathcal W(q_0)&=\frac{\mathcal W_D^{\star}(q_0)/N-\mathcal W_R^{\star}(q_0)}{\hbar\omega_0}.
\end{align}
Positive values favor the common-cavity Dicke battery under the stated input bound.

At $q_0=0.5$, Fig.~\ref{fig:dicke-independent-equal-input}(a) and Fig.~\ref{fig:dicke-independent-equal-input}(b) show positive stored-energy and ergotropy differences for every displayed $N>1$ and pump strength.
The differences are largest at $\gamma/\omega_0=0.1$, for which the weaker pump takes longer to exhaust the same input budget and leaves more time for coherent transfer from the common cavity.
This result describes the use of a limited pump input and does not imply a larger charging power.

At $q_0=1$, the ergotropy difference at $\gamma/\omega_0=0.1$ remains positive over the displayed range, whereas the energy difference at the same pump strength becomes slightly negative at $N=20$.
For $\gamma/\omega_0=0.5$ and $1$, both differences are negative for every $N>1$.
The differences at $\gamma/\omega_0=0.1$ grow from the $N=1$ baseline, reach their largest values near $N=8$--$10$, and then decrease.
The relative performance therefore depends on the input bound, pump strength, observable, and battery size.

At the displayed bound $q_0=0.5$, the common-cavity battery gives the larger budget-constrained stored energy and ergotropy, whereas the ordering at $q_0=1$ depends on the pump strength, observable, and battery size.
The larger values at $\gamma/\omega_0=0.1$ and $q_0=0.5$ are consistent with the longer time available for common-cavity transfer before the input bound is first reached; coherent backflow and pump-induced mixedness subsequently separate the stored-energy and ergotropy trends.
These two input bounds establish a resource-dependent ordering but do not locate a universal crossover or demonstrate a resource-independent collective advantage.

\section{Conclusions}
\label{sec:conclusion}
We have studied open charging protocols for Rabi and Dicke quantum batteries under direct incoherent pumping.
For the Rabi battery, pumping suppresses coherent energy backflow, broadens the high-energy charging region, and reduces the sensitivity to both the stopping time and the detuning.
The unconditional battery-current balance shows that the positive pump current offsets the negative cavity current during coherent return.
In the conditioned no-jump picture, the same pumping favors higher battery-excitation sectors in both architectures.
For the Rabi battery, explicit trajectory weights at $\gamma/\omega_0=1$ show that close reduced-battery observables can coexist with a small no-jump probability because the no-jump and jump-conditioned sectors approach similar highly excited battery states.
In a separate Rabi operating-point comparison, direct incoherent pumping yields larger extractable work and a larger input-bookkeeping yield than the Hermitian reference under the adopted bookkeeping convention.

The finite-size scan shows observable-dependent tradeoffs at $\gamma/\omega_0=0.1$ and $0.5$, whereas at $\gamma/\omega_0=1$ local incoherent pumping improves the peak stored energy, peak ergotropy, and near-peak timing robustness relative to the Hermitian reference throughout the scanned size range.
The maximum average charging power increases with the pump strength, although the energy-flow decomposition shows that $\Delta P_{\max}$ is pump dominated at $\gamma/\omega_0=0.5$ and $1$.
Under equal accumulated pump input, the common-cavity Dicke battery yields larger budget-constrained stored energy and ergotropy at the displayed bound $q_0=0.5$, whereas the ordering at $q_0=1$ depends on the pump strength, observable, and battery size.
The two tested input bounds therefore demonstrate resource-dependent architecture ordering without locating a universal crossover or establishing a resource-independent collective advantage.

Taken together, these results show that direct incoherent pumping provides a route to stabilized charging in the Rabi and finite Dicke quantum batteries considered here.
The central experimental requirement is an incoherent excitation channel that acts predominantly on the TLS battery while preserving the cavity field over the charging interval.
In circuit QED, this requires spectrally and spatially selective TLS pumping and a cavity linewidth below the inverse charging time \cite{naghilooQuantum2019a, liObservation2019a, PhysRevA.108.053712}.
Cavity-QED arrays and trapped-ion systems also offer controlled access to collective spin-boson interactions and localized losses \cite{baumannDicke2010, Hu_2022, PhysRevA.103.L020201}.
Cavity loss and spontaneous emission are omitted here to isolate the pumping mechanism.
These channels are expected to lower the stored energy and ergotropy, shorten the high-energy plateau, and shift the optimal pump window; quantifying this competition requires an extended dissipative model.

\begin{acknowledgments}
This work was supported by the National Natural Science Foundation of China (Grant Nos. 12205383 and 12274470) and the Hunan Provincial Natural Science Foundation (Grant No. 2024JJ6483). 
A.-L. Zhou was also supported by the Fundamental Research Funds for the Central Universities of Central South University under Grant No. 1053320214335. 
\end{acknowledgments}

\appendix

\section{Conditioned no-jump derivations and Tavis--Cummings reduction} \label{DENHH}

The Lindblad dynamics of the Rabi and Dicke batteries is defined in Sec.~\ref{sec:model} \cite{Ashida02072020, huber_emergence_2020, dast_quantum_2014, ManzanoAIP, nakanishi_pt_2022}.
Within the quantum trajectory framework, post-selecting the continuous null-measurement record gives an unnormalized state generated by
\begin{align}
    H_X^{\rm eff}=H_X-\frac{i}{2}\sum_kL_k^\dagger L_k,
    \qquad X={\rm R,D},
    \label{eq:general-effective-generator}
\end{align}
where $H_X$ denotes $H_{\rm R}$ or $H_{\rm D}$ and the sum includes the jump operators of the corresponding model.

\subsection{Rabi battery derivation}

For the physical Rabi protocol, the pump jump operator is $L=\sqrt{2\gamma}\sigma_+$.
The corresponding dissipative term evaluates to $L^\dagger L=2\gamma|g\rangle\langle g|$.
Using the Pauli identity $|g\rangle\langle g| = (\mathbb{I} - \sigma_z)/2$ in Eq.~(\ref{eq:general-effective-generator}) gives the Rabi generator in Eq.~(\ref{eq:rabi-effective-generator}).
The global decay term $-i\gamma\mathbb I/2$ induces a uniform norm decay and is removed by the instantaneous state normalization $|\psi(t)\rangle \to |\psi(t)\rangle / \| \psi(t) \|$ used to evaluate conditioned observables.
The remaining term introduces a relative gain for the excited state.
Thus the nontrivial conditioned bias is $i\gamma\sigma_z/2$, while the full generator $H_{\rm R}^{\rm eff}$ must be retained when its norm is used as the no-jump probability.

\subsection{Dicke battery derivation under identical local pumping}

For the Dicke Lindblad benchmark, each TLS battery is pumped locally with
\begin{align}
    L_k=\sqrt{2\gamma}\sigma_+^{(k)},\qquad k=1,\ldots,N .
\end{align}
The no-jump effective Hamiltonian is therefore
\begin{align}
    H_{\rm D}^{\rm eff}
    &=
    H_{\rm D}
    -
    \frac{i}{2}\sum_{k=1}^{N}L_k^\dagger L_k  \notag\\
    &=
    H_{\rm D}-i\gamma
    \sum_{k=1}^{N}\sigma_-^{(k)}\sigma_+^{(k)} .
\end{align}
Using $\sigma_-\sigma_+=|g\rangle\langle g|=(\mathbb{I}-\sigma_z)/2$, one obtains
\begin{align}
    \sum_{k=1}^{N}\sigma_-^{(k)}\sigma_+^{(k)}
    =
    \frac{N}{2}-J_z .
\end{align}
Substitution into Eq.~(\ref{eq:general-effective-generator}) gives the Dicke generator in Eq.~(\ref{eq:dicke-effective-generator}).
The final term is a state-independent imaginary shift and is removed by the same instantaneous normalization used for the no-jump trajectory.
The remaining term $i\gamma J_z$ produces the conditioned Dicke bias discussed in Sec.~\ref{sec:dicke-nh-mechanism}.
This derivation starts from identical local Lindblad pumps and keeps the local dissipator distinct from any collective-channel approximation.

\subsection{Tavis--Cummings reduction for the Dicke battery}

To expose the collective scale used in Sec.~\ref{sec:dicke-nh-mechanism}, we remove the same state-independent shift, apply the rotating-wave approximation, and use $J_\pm=J_x\pm iJ_y$ in the effective Tavis--Cummings generator
\begin{align}
    H_{\rm TC}^{\rm eff}
    & =
    \omega_c a^\dagger a
    +
    \omega_0 J_z
    +
    g\left(a^\dagger J_- + a J_+\right)
    +
    i\gamma J_z .
\end{align}
Within the fully symmetric sector of $N$ TLSs, this auxiliary model conserves
\begin{align}
    M = a^\dagger a + J_z + \frac{N}{2}
\end{align}
and decomposes into fixed-$M$ blocks.
In the single-excitation block, the resulting two-level sector yields the collective complex frequency
\begin{align}
    \widetilde{\Omega}_N
    =
    \sqrt{4g^2N+(\Delta+i\gamma)^2},
\end{align}
which makes the collective scale $g\sqrt{N}$ and the auxiliary time scale $t_{\rm col}\sim1/(g\sqrt{N})$ explicit.
In the corresponding single-excitation reduction, $N=1$ recovers the Rabi result $\widetilde{\Omega}_1$.
This reduction organizes only the conditioned no-jump dynamics, and no $N$-scaling law for the Lindblad stored energy or charging power is inferred from it.

\section{Numerical details and conditional-sector diagnostic}
\label{app:numerical-controls}
Unless otherwise stated in the corresponding caption, the Rabi simulations use $\omega_0=1$, $\Delta=\omega_0-\omega_c$, the initial state $|\psi(0)\rangle=|1\rangle_c\otimes|g\rangle_B$, and jump operator $L=\sqrt{2\gamma}\sigma_+$.
Because the Dicke benchmark uses identical local pumps $L_k=\sqrt{2\gamma}\sigma_+^{(k)}$ and identical cavity couplings for all TLSs, its Liouvillian is invariant under permutations of the TLS labels.
We propagate the Lindblad density operator in the permutation-invariant Dicke-basis representation of Ref.~\cite{PhysRevA.98.063815}, which retains the total-spin sectors and their degeneracies that can be populated by local pump jumps.
The Hermitian and conditioned no-jump Dicke branches preserve the symmetric $J=N/2$ sector for the cavity-loaded initial state and are evaluated in that sector.
The minimal open-system model includes the pumping channel only; cavity loss and spontaneous emission are omitted in order to isolate the role of direct incoherent pumping and its associated gain-loss bias.

\begin{figure}[t]
    \centering
    \includegraphics[width=\linewidth]{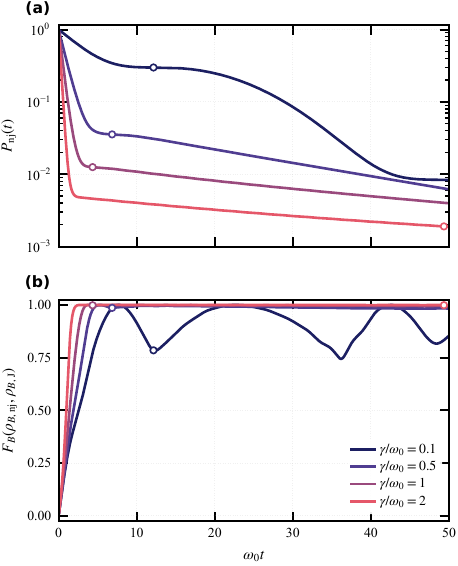}
    \caption{Statistical weight and conditional-state comparison for the Rabi battery at $g/\omega_0=0.1$, $\Delta=0$, $n_{\rm ph}=10$, and jump operator $L=\sqrt{2\gamma}\sigma_+$.
    (a) Probability $P_{\rm nj}(t)$ that no pump jump has occurred up to time $t$.
    (b) Uhlmann root fidelity between the reduced battery states conditioned on no jump and on at least one pump jump.
    The fidelity is shown after the jump-conditioned sector acquires nonzero weight.
    Open circles mark the energy maximum of the normalized no-jump trajectory.
    Increasing $\gamma$ lowers the no-jump probability while bringing the two conditional reduced battery states closer to one another.}
    \label{fig:nojump-jump-sectors}
\end{figure}

For the upper row representing the Hermitian reference in Fig.~\ref{fig:single-battery-benchmarks}, the coupling scan uses the photon-number cutoff $n_{\mathrm{ph}}=14$ after convergence checks with larger cutoffs.
For the full-Lindblad lower row, we use $n_{\mathrm{ph}}=24$ and a 1001-point time grid over $\omega_0t\in[0,50]$; panel (e) displays the interval $0\leq\omega_0t\leq30$, and panels (f)--(h) use all 41 calculated values $\gamma/\omega_0=0,0.05,\ldots,2$.
For Fig.~\ref{fig:single-battery-applicability}, the benchmark-based usage maps use $100\times100$ parameter grids, $n_{\mathrm{ph}}=10$, and a uniform time grid $\omega_0t_j\in[0,50]$ with $N_t=201$ points.
The bottom row of Fig.~\ref{fig:single-battery-applicability} averages the reduced-battery-state fidelity and measures battery-subsystem agreement over the simulation window.
For the conditional-sector diagnostic in Fig.~\ref{fig:nojump-jump-sectors}, we use $g/\omega_0=0.1$, $\Delta=0$, $n_{\rm ph}=10$, and a uniform grid $\omega_0t\in[0,50]$ with 1001 points.
The no-jump probability is evaluated from $H_{\rm R}^{\rm eff}$ before the state-independent term $-i\gamma\mathbb{I}/2$ is removed, while the normalized no-jump battery state is unchanged by this term.
At the normalized no-jump energy maximum, the pair $(P_{\rm nj},F_B)$ changes from $(0.2988,0.7849)$ at $\gamma/\omega_0=0.1$ to $(0.0125,0.9985)$ at $\gamma/\omega_0=1$, where $F_B$ is the Uhlmann root fidelity between the reduced battery states conditioned on no jump and on at least one jump.

The Dicke benchmarks use the cavity-loaded initial condition described in Sec.~\ref{sec:model}.
For Fig.~\ref{fig:dicke-direct-pumping-response}, the fixed-$N$ Dicke benchmark uses $N=10$, $g/\omega_0=0.1$, photon cutoff $n_{\mathrm{ph}}=26$, and a uniform time grid $\omega_0t\in[0,40]$ with 601 points.
Repeating the calculation with $n_{\rm ph}=20$, $26$, and $32$ for $\gamma/\omega_0=0$, $0.5$, and $1$ gives a maximum relative change of $1.63\times10^{-6}$ in $\Delta E_{\max}/(N\hbar\omega_0)$ between the two largest cutoffs.
Panels (b)--(d) use the pump range $0\le\gamma/\omega_0\le2$.
The representative dynamics in panel (a) show selected pump strengths from the same data set.
For Fig.~\ref{fig:dicke-finite-size}, the calculations use $N=4,6,\ldots,20$, $\gamma/\omega_0=\{0,0.1,0.5,1\}$, $n_{\mathrm{ph}}=\max(24,N+16)$, and 601 time points over $\omega_0t\in[0,40]$.
The reported maxima and threshold windows are evaluated over this interval.
For the pumped Dicke power, the reported maximum is the larger of the sampled finite-time value and the analytical initial limit $\Delta P/(N\hbar\omega_0^2)=2\gamma/\omega_0$.
The Dicke ergotropy is evaluated from the reduced battery state using a passive-state construction that retains the degeneracies of the Dicke sectors represented in the permutation-invariant basis.
At the ergotropy-optimal time $t_{\mathcal W}^{*}$, we also evaluate the extractable fraction $f_{\mathcal W}=\mathcal W(t_{\mathcal W}^{*})/\Delta E(t_{\mathcal W}^{*})$.
Across this grid, the fraction spans $0.893$--$0.938$, $0.670$--$0.774$, $0.839$--$0.892$, and $0.925$--$0.956$ for $\gamma/\omega_0=0$, $0.1$, $0.5$, and $1$, respectively.
For Fig.~\ref{fig:dicke-independent-equal-input}, both architectures use $g/\omega_0=0.1$, the same total initial photon energy, and a 601-point time grid over $\omega_0t\in[0,40]$.
The Dicke cutoff is $n_{\rm ph}=\max(24,N+16)$, the independent Rabi calculation uses $n_{\rm ph}=24$, and the displayed comparisons use $N=1,2,4,6,\ldots,20$, $\gamma/\omega_0=0.1,0.5,1$, and $q_0=0.5,1$.

\section{Energy-flow decomposition and exchange dynamics}
\label{app:dicke-energy-flow}

For the locally pumped Dicke battery, let $E_B(t)=\operatorname{Tr}[H_B\rho_B(t)]$ denote the battery energy.
With the interaction Hamiltonian $H_{\rm int}=2g(a^\dagger+a)J_x$, its rate of change separates as
\begin{align}
\frac{dE_B}{dt}=J_{\rm cav}(t)+J_{\rm pump}(t),
\end{align}
where
\begin{align}
J_{\rm cav}(t)
&=-i\operatorname{Tr}\!\left\{H_B[H_{\rm int},\rho(t)]\right\},\\
J_{\rm pump}(t)
&=2\gamma\sum_{k=1}^{N}
\operatorname{Tr}\!\left\{H_B\mathcal D[\sigma_+^{(k)}]\rho(t)\right\}.
\end{align}
In these traces, $H_B$ denotes the battery Hamiltonian tensored with the identity operator of the cavity.
The corresponding accumulated energies are $Q_x(t)=\int_0^tJ_x(t')dt'$ for $x\in\{\mathrm{cav},\mathrm{pump}\}$, giving
\begin{align}
\Delta P(t)
=\frac{\Delta E(t)}{t}
=\frac{Q_{\rm cav}(t)}{t}+\frac{Q_{\rm pump}(t)}{t}.
\end{align}

\begin{figure}[t]
    \centering
    \includegraphics[width=0.94\columnwidth]{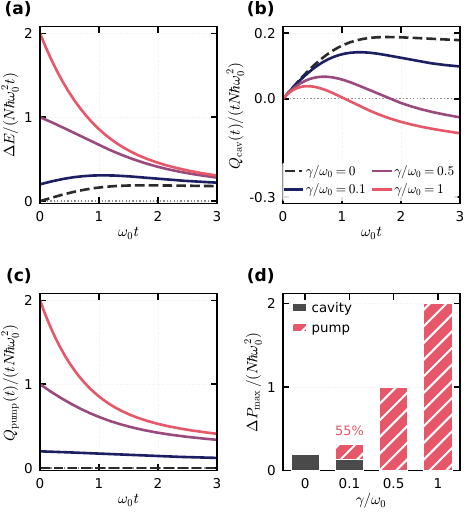}
    \caption{Energy-flow decomposition of the average charging power for the locally pumped Dicke battery at $N=10$, $g/\omega_0=0.1$, and $n_{\rm ph}=26$.
    Panels (a)--(c) show the total average power, the accumulated cavity contribution $Q_{\rm cav}(t)/t$, and the accumulated pump contribution $Q_{\rm pump}(t)/t$, respectively.
    Black dashed, dark-indigo, purple, and red curves correspond to $\gamma/\omega_0=0,0.1,0.5,1$.
    Panel (d) separates the cavity and pump contributions at the time when the total average charging power reaches $\Delta P_{\max}$.
    The label $55\%$ gives the pump fraction at $\gamma/\omega_0=0.1$.}
    \label{fig:dicke-energy-flow}
\end{figure}

Figure~\ref{fig:dicke-energy-flow} shows this decomposition for the representative $N=10$ Dicke battery.
For the Hermitian reference, the maximum average charging power is $\Delta P_{\max}/(N\hbar\omega_0^2)=0.188$ at $\omega_0t=1.772$ and comes from the cavity.
At $\gamma/\omega_0=0.1$, the maximum is $0.307$ at $\omega_0t=1.068$; the cavity and pump contributions are $0.138$ and $0.169$, respectively.
The pump therefore supplies $55\%$ of this finite-time maximum, while the cavity supplies $45\%$.

For the initial state $|N\rangle_c\otimes|j,-j\rangle_B$, the cavity contribution vanishes at the initial time and the pump gives
\begin{align}
\lim_{t\rightarrow0^+}
\frac{\Delta E(t)}{N\hbar\omega_0^2t}
=2\frac{\gamma}{\omega_0}.
\end{align}
This initial limit sets the maximum average charging power at $\gamma/\omega_0=0.5$ and $1$, yielding the values $1$ and $2$ in Fig.~\ref{fig:dicke-energy-flow}(d).
The accumulated cavity contribution $Q_{\rm cav}(t)/t$ in Fig.~\ref{fig:dicke-energy-flow}(b) becomes negative for $\gamma/\omega_0=0.5$ and $1$, indicating a net transfer out of the battery through the interaction channel over the elapsed interval.
Its instantaneous direction is determined by $J_{\rm cav}(t)$.

To resolve how pumping acts on this exchange, define $X=a+a^\dagger$, $Y=i(a^\dagger-a)$, and the joint correlation $C(t)=2\langle XJ_y\rangle/N$.
For the full Hamiltonian and local jumps used here, the adjoint Lindblad equation gives the exact relations
\begin{align}
J_{\rm cav}(t)&=Ng\omega_0 C(t),\label{eq:exchange-current}\\
\dot C(t)&=S_H(t)-\gamma C(t),\label{eq:exchange-dynamics}\\
S_H(t)&=\frac{2\omega_c}{N}\langle YJ_y\rangle
+\frac{2\omega_0}{N}\langle XJ_x\rangle\nonumber\\
&\quad-\frac{4g}{N}\langle X^2J_z\rangle.\label{eq:exchange-source}
\end{align}
For $N=1$ and $J_\mu=\sigma_\mu/2$, they also describe the Rabi battery, with $C=\langle X\sigma_y\rangle$.

The local pumps contribute the damping term $-\gamma C$ and also modify the Hamiltonian source $S_H$ through the evolving state.
The cavity current is therefore governed by the balance between these two terms.
In the battery-energy balance, the positive pump current reduces net discharge even while the cavity current remains negative.

\section{Analytical solutions and exceptional-point dynamics} \label{ASEPD}
\begin{figure}[tb]
    \centering
    \includegraphics[width=\linewidth]{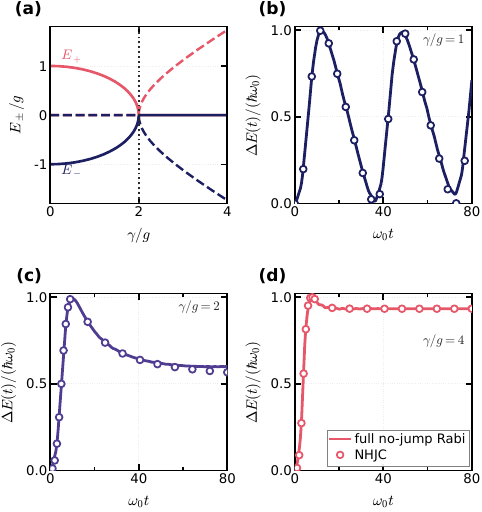}
    \caption{Spectral and charging dynamics of the conditioned models at $g/\omega_0=0.1$ and exact resonance ($\Delta=0$).
    Panel (a) shows $\operatorname{Re}E_\pm/g$ as solid curves and $\operatorname{Im}E_\pm/g$ as dashed curves versus $\gamma/g$; red and dark indigo denote the $E_+$ and $E_-$ branches, and the dotted line marks the exceptional point $\gamma=2g$.
    Panels (b)--(d) compare the full normalized no-jump Rabi numerics (solid curves) with the analytical non-Hermitian Jaynes--Cummings result (open circles) derived in Appendix~\ref{ASEPD}.
    Within each panel the numerical and analytical results share a color; dark indigo, purple, and red encode $\gamma/g=1$, $2$, and $4$, corresponding to the weak-pumping, exceptional-point, and strong-pumping regimes of the resonant conditioned NHJC model.
    The stored energy in panels (b)--(d) is normalized by $\hbar\omega_0$.}
    \label{Appendix_B}
\end{figure}

We focus on the weak-coupling regime in which $g \ll \omega_0, \omega_c$.
In this limit, the counter-rotating terms in the quantum Rabi model oscillate rapidly and can be neglected over the relevant timescales.
We subsequently apply the rotating-wave approximation (RWA).
This approximation maps the dynamics onto an effective non-Hermitian Jaynes--Cummings (NHJC) model.
In this appendix, we derive the analytical solutions for this effective model.
These results are used as mechanism-level guidance for the weak-coupling regime and are compared with the full no-jump Rabi numerics in Fig.~\ref{Appendix_B}.

After removing the state-independent term $-i\gamma\mathbb I/2$, which does not affect instantaneously normalized observables, the effective Jaynes--Cummings generator is
\begin{align}
    H_{\rm JC}^{\rm eff} = \omega_c a^\dagger a + \frac{\omega_0 + i\gamma}{2}\sigma_z + g(\sigma_- a^\dagger + \sigma_+ a).
\end{align}
The remaining term $i\gamma\sigma_z/2$ biases the excited component relative to the ground component and makes the auxiliary evolution nonunitary.

Assuming the system is initialized in $|\psi(0)\rangle=|1\rangle_c\otimes|g\rangle_B$, the dynamics remain within the single-excitation subspace spanned by the ordered basis $\mathcal{B}=\{|0\rangle_c\otimes|e\rangle_B,|1\rangle_c\otimes|g\rangle_B\}$.
After restricting $H_{\rm JC}^{\rm eff}$ to this subspace and extracting a global average-energy phase, its nontrivial block is
\begin{align}
    \frac{1}{2} \begin{pmatrix} \widetilde{\Delta} & \Omega_R \\ \Omega_R & -\widetilde{\Delta} \end{pmatrix},
\end{align}
where we have defined the complex detuning $\widetilde{\Delta} = \Delta + i\gamma$ (with the standard detuning $\Delta = \omega_0 - \omega_c$) and the Rabi frequency $\Omega_R = 2g$.

The eigenvalues of this block analytically govern the transient dynamics.
Defining the generalized complex Rabi frequency
\begin{align}
    \widetilde{\Omega}_1 = \sqrt{\widetilde{\Delta}^2 + \Omega_R^2} = \sqrt{(\Delta + i\gamma)^2 + 4g^2}, \label{Omega}
\end{align}
the eigenenergies read
\begin{align}
    E_{\pm} = \pm \frac{1}{2} \widetilde{\Omega}_1 = \pm \frac{1}{2} \sqrt{\widetilde{\Delta}^2 + \Omega_R^2}.
\end{align}
At exact resonance ($\Delta=0$), the spectrum reduces to $E_{\pm} = \pm \frac{1}{2} \sqrt{4g^2 - \gamma^2}$ [see Fig.~\ref{Appendix_B}(a)].
Within the resonant conditioned NHJC model, the exceptional point (EP) occurs at $\gamma=2g$.
We call $\gamma<2g$ the weak-pumping regime and $\gamma>2g$ the strong-pumping regime; the eigenenergies are real and purely imaginary, respectively.

The corresponding phase-reduced time-evolution operator $U(t)$ is exactly
\begin{equation}
    U(t) =
    \begin{pmatrix}
        c-i(\widetilde{\Delta}/\widetilde{\Omega}_1)s & -i(\Omega_R/\widetilde{\Omega}_1)s \\
        -i(\Omega_R/\widetilde{\Omega}_1)s & c+i(\widetilde{\Delta}/\widetilde{\Omega}_1)s
    \end{pmatrix},
\end{equation}
where
\begin{equation*}
    c = \cos\left(\frac{\widetilde{\Omega}_1t}{2}\right),
    \qquad
    s = \sin\left(\frac{\widetilde{\Omega}_1t}{2}\right).
\end{equation*}

Applying $U(t)$ to the initial state $|\psi(0)\rangle = (0, 1)^T$, we obtain the generally unnormalized auxiliary amplitudes $C_e(t)$ and $C_g(t)$ for the composite basis states $|0\rangle_c\otimes|e\rangle_B$ and $|1\rangle_c\otimes|g\rangle_B$, respectively:
\begin{align}
    C_e(t) &= -i \frac{\Omega_R}{\widetilde{\Omega}_1} \sin\left(\frac{\widetilde{\Omega}_1t}{2}\right),  \\
    C_g(t) &= \cos\left(\frac{\widetilde{\Omega}_1t}{2}\right) + i \frac{\widetilde{\Delta}}{\widetilde{\Omega}_1} \sin\left(\frac{\widetilde{\Omega}_1t}{2}\right).
\end{align}
The norm of this shift-removed auxiliary state is not the physical no-jump probability, which is obtained from the complete generator before the identity term is removed.
The conditioned population is therefore evaluated from the instantaneously normalized amplitudes.
The normalized excited-state population is
\begin{align}
    \bar{P}_e(t) = \frac{|C_e(t)|^2}{|C_e(t)|^2 + |C_g(t)|^2}.
\end{align}
According to the definition of the net charging energy $\Delta E(t) = \text{Tr}[H_B \rho_B(t)] - \text{Tr}[H_B \rho_B(0)]$ with $H_B = \frac{\omega_0}{2}\sigma_z$, the time-dependent charging energy is proportional to the excited-state population.
It is therefore $\Delta E(t) = \omega_0 \bar{P}_e(t)$.
Using the abbreviations $c=\cos(\widetilde{\Omega}_1t/2)$ and $s=\sin(\widetilde{\Omega}_1t/2)$ introduced with $U(t)$, substitution of these amplitudes gives
\begin{align}
    \Delta E(t)
    =\frac{\omega_0\left|(\Omega_R/\widetilde{\Omega}_1)s\right|^2}
    {\left|(\Omega_R/\widetilde{\Omega}_1)s\right|^2
    +\left|c+i(\widetilde{\Delta}/\widetilde{\Omega}_1)s\right|^2}. \label{deltaE}
\end{align}

This general solution captures the critical damping mechanism at the EP.
At $\gamma = 2g$, the eigenvalues coalesce ($E_{\pm}=0$) and the complex Rabi frequency vanishes ($\widetilde{\Omega}_1 = 0$).
Taking the limit $\widetilde{\Omega}_1 \to 0$, the trigonometric evolution degenerates into a polynomial time dependence:
\begin{align}
    \Delta E_{\text{EP}}(t) = \omega_0 \frac{g^2 t^2}{1 - 2gt + 2g^2 t^2}.
\end{align}
As confirmed by the numerical simulations at the EP in Fig.~\ref{Appendix_B}(c), this polynomial trajectory embodies a critically damped dynamical phase.
At this point, the balance between the pump-induced non-Hermitian bias and coherent exchange removes the oscillatory backflow that characterizes the underdamped regime, thereby marking the transition to an oscillation-free charging regime within the effective model.

\section{Asymptotic limit of the effective dynamics and frequency robustness}\label{ENHH}
Having established the transient dynamics in Appendix \ref{ASEPD}, we now examine the long-time asymptotic behavior of the effective dynamics.
The following asymptotic argument applies in parameter regimes where the complex Rabi frequency is chosen with $\operatorname{Im}\widetilde{\Omega}_1>0$.
This includes the strong-pumping regime of the resonant conditioned NHJC model and detuned cuts for which $\operatorname{Im}\widetilde{\Omega}_1>0$ controls the late-time trigonometric factors.

When $\operatorname{Im}\widetilde{\Omega}_1>0$, the normalized effective dynamics approaches a saturated late-time value.
This value describes the plateau of the corresponding conditioned dynamics.
Taking the corresponding limit $t \to \infty$, the ratio of the trigonometric functions converges to a constant:
\begin{align}
    \cos(\widetilde{\Omega}_1t/2) / \sin(\widetilde{\Omega}_1t/2) \to -i.
\end{align}
Consequently, the ratio of the unnormalized amplitudes converges to
\begin{align}
    \lim_{t \to \infty} \frac{C_g(t)}{C_e(t)} = \frac{\widetilde{\Omega}_1 - \widetilde{\Delta}}{\Omega_R}.
\end{align}
Substituting this asymptotic limit into the instantaneous normalization formula yields the analytical expression for the asymptotic stored energy:
\begin{align}
    E_{\mathrm{asym}} =  \frac{\omega_0 |\Omega_R|^2}{|\Omega_R|^2 + |\widetilde{\Omega}_1 - \widetilde{\Delta}|^2}. \label{Emax-delta}
\end{align}
This analytical expression accounts for the detuning profile of the strong-pumping plateau within the effective Jaynes--Cummings description.
In the relevant monotonic windows, it explains the frequency-robust plateau observed in the corresponding conditioned full-Rabi simulations.
In the asymptotic limit $\gamma \gg g,|\Delta|$, the non-Hermitian contribution dominates the detuning dependence in the asymptotic expression.
As a result, the effective dynamics becomes much less sensitive to the real detuning and develops a broader high-energy region in parameter space.
This asymptotic expression is not an analytical solution of the physical Lindblad dynamics, whose detuning response remains established by the master-equation calculation in Fig.~\ref{fig:detuning-robustness}.


\begin{thebibliography}{53}%
\makeatletter
\providecommand \@ifxundefined [1]{%
 \@ifx{#1\undefined}
}%
\providecommand \@ifnum [1]{%
 \ifnum #1\expandafter \@firstoftwo
 \else \expandafter \@secondoftwo
 \fi
}%
\providecommand \@ifx [1]{%
 \ifx #1\expandafter \@firstoftwo
 \else \expandafter \@secondoftwo
 \fi
}%
\providecommand \natexlab [1]{#1}%
\providecommand \enquote  [1]{``#1''}%
\providecommand \bibnamefont  [1]{#1}%
\providecommand \bibfnamefont [1]{#1}%
\providecommand \citenamefont [1]{#1}%
\providecommand \href@noop [0]{\@secondoftwo}%
\providecommand \href [0]{\begingroup \@sanitize@url \@href}%
\providecommand \@href[1]{\@@startlink{#1}\@@href}%
\providecommand \@@href[1]{\endgroup#1\@@endlink}%
\providecommand \@sanitize@url [0]{\catcode `\\12\catcode `\$12\catcode
  `\&12\catcode `\#12\catcode `\^12\catcode `\_12\catcode `\%12\relax}%
\providecommand \@@startlink[1]{}%
\providecommand \@@endlink[0]{}%
\providecommand \url  [0]{\begingroup\@sanitize@url \@url }%
\providecommand \@url [1]{\endgroup\@href {#1}{\urlprefix }}%
\providecommand \urlprefix  [0]{URL }%
\providecommand \Eprint [0]{\href }%
\providecommand \doibase [0]{https://doi.org/}%
\providecommand \selectlanguage [0]{\@gobble}%
\providecommand \bibinfo  [0]{\@secondoftwo}%
\providecommand \bibfield  [0]{\@secondoftwo}%
\providecommand \translation [1]{[#1]}%
\providecommand \BibitemOpen [0]{}%
\providecommand \bibitemStop [0]{}%
\providecommand \bibitemNoStop [0]{.\EOS\space}%
\providecommand \EOS [0]{\spacefactor3000\relax}%
\providecommand \BibitemShut  [1]{\csname bibitem#1\endcsname}%
\let\auto@bib@innerbib\@empty
\bibitem [{\citenamefont {Alicki}\ and\ \citenamefont
  {Fannes}(2013)}]{alicki_entanglement_2013}%
  \BibitemOpen
  \bibfield  {author} {\bibinfo {author} {\bibfnamefont {R.}~\bibnamefont
  {Alicki}}\ and\ \bibinfo {author} {\bibfnamefont {M.}~\bibnamefont
  {Fannes}},\ }\bibfield  {title} {\bibinfo {title} {Entanglement boost for
  extractable work from ensembles of quantum batteries},\ }\href
  {https://doi.org/10.1103/PhysRevE.87.042123} {\bibfield  {journal} {\bibinfo
  {journal} {Phys. Rev. E}\ }\textbf {\bibinfo {volume} {87}},\ \bibinfo
  {pages} {042123} (\bibinfo {year} {2013})}\BibitemShut {NoStop}%
\bibitem [{\citenamefont {Binder}\ \emph {et~al.}(2015)\citenamefont {Binder},
  \citenamefont {Vinjanampathy}, \citenamefont {Modi},\ and\ \citenamefont
  {Goold}}]{Felix_Quantacell_2015}%
  \BibitemOpen
  \bibfield  {author} {\bibinfo {author} {\bibfnamefont {F.~C.}\ \bibnamefont
  {Binder}}, \bibinfo {author} {\bibfnamefont {S.}~\bibnamefont
  {Vinjanampathy}}, \bibinfo {author} {\bibfnamefont {K.}~\bibnamefont
  {Modi}},\ and\ \bibinfo {author} {\bibfnamefont {J.}~\bibnamefont {Goold}},\
  }\bibfield  {title} {\bibinfo {title} {Quantacell: powerful charging of
  quantum batteries},\ }\href {https://doi.org/10.1088/1367-2630/17/7/075015}
  {\bibfield  {journal} {\bibinfo  {journal} {New J. Phys.}\ }\textbf {\bibinfo
  {volume} {17}},\ \bibinfo {pages} {075015} (\bibinfo {year}
  {2015})}\BibitemShut {NoStop}%
\bibitem [{\citenamefont {Campaioli}\ \emph {et~al.}(2017)\citenamefont
  {Campaioli}, \citenamefont {Pollock}, \citenamefont {Binder}, \citenamefont
  {C\'eleri}, \citenamefont {Goold}, \citenamefont {Vinjanampathy},\ and\
  \citenamefont {Modi}}]{Campaioli_Enhancing_2018}%
  \BibitemOpen
  \bibfield  {author} {\bibinfo {author} {\bibfnamefont {F.}~\bibnamefont
  {Campaioli}}, \bibinfo {author} {\bibfnamefont {F.~A.}\ \bibnamefont
  {Pollock}}, \bibinfo {author} {\bibfnamefont {F.~C.}\ \bibnamefont {Binder}},
  \bibinfo {author} {\bibfnamefont {L.}~\bibnamefont {C\'eleri}}, \bibinfo
  {author} {\bibfnamefont {J.}~\bibnamefont {Goold}}, \bibinfo {author}
  {\bibfnamefont {S.}~\bibnamefont {Vinjanampathy}},\ and\ \bibinfo {author}
  {\bibfnamefont {K.}~\bibnamefont {Modi}},\ }\bibfield  {title} {\bibinfo
  {title} {Enhancing the charging power of quantum batteries},\ }\href
  {https://doi.org/10.1103/PhysRevLett.118.150601} {\bibfield  {journal}
  {\bibinfo  {journal} {Phys. Rev. Lett.}\ }\textbf {\bibinfo {volume} {118}},\
  \bibinfo {pages} {150601} (\bibinfo {year} {2017})}\BibitemShut {NoStop}%
\bibitem [{\citenamefont {Campaioli}\ \emph {et~al.}(2024)\citenamefont
  {Campaioli}, \citenamefont {Gherardini}, \citenamefont {Quach}, \citenamefont
  {Polini},\ and\ \citenamefont {Andolina}}]{RevModPhys.96.031001}%
  \BibitemOpen
  \bibfield  {author} {\bibinfo {author} {\bibfnamefont {F.}~\bibnamefont
  {Campaioli}}, \bibinfo {author} {\bibfnamefont {S.}~\bibnamefont
  {Gherardini}}, \bibinfo {author} {\bibfnamefont {J.~Q.}\ \bibnamefont
  {Quach}}, \bibinfo {author} {\bibfnamefont {M.}~\bibnamefont {Polini}},\ and\
  \bibinfo {author} {\bibfnamefont {G.~M.}\ \bibnamefont {Andolina}},\
  }\bibfield  {title} {\bibinfo {title} {Colloquium: Quantum batteries},\
  }\href {https://doi.org/10.1103/RevModPhys.96.031001} {\bibfield  {journal}
  {\bibinfo  {journal} {Rev. Mod. Phys.}\ }\textbf {\bibinfo {volume} {96}},\
  \bibinfo {pages} {031001} (\bibinfo {year} {2024})}\BibitemShut {NoStop}%
\bibitem [{\citenamefont {Li}\ \emph {et~al.}(2025)\citenamefont {Li},
  \citenamefont {He}, \citenamefont {Ouyang}, \citenamefont {Feng},\ and\
  \citenamefont {Wei}}]{li_switchable_2025}%
  \BibitemOpen
  \bibfield  {author} {\bibinfo {author} {\bibfnamefont {Y.~F.}\ \bibnamefont
  {Li}}, \bibinfo {author} {\bibfnamefont {S.~R.}\ \bibnamefont {He}}, \bibinfo
  {author} {\bibfnamefont {P.~H.}\ \bibnamefont {Ouyang}}, \bibinfo {author}
  {\bibfnamefont {X.~N.}\ \bibnamefont {Feng}},\ and\ \bibinfo {author}
  {\bibfnamefont {L.~F.}\ \bibnamefont {Wei}},\ }\bibfield  {title} {\bibinfo
  {title} {Switchable coherent quantum batteries with large ergotropy},\ }\href
  {https://doi.org/10.1103/jpcs-998q} {\bibfield  {journal} {\bibinfo
  {journal} {Phys. Rev. A}\ }\textbf {\bibinfo {volume} {112}},\ \bibinfo
  {pages} {022217} (\bibinfo {year} {2025})}\BibitemShut {NoStop}%
\bibitem [{\citenamefont {Shi}\ \emph {et~al.}(2022)\citenamefont {Shi},
  \citenamefont {Ding}, \citenamefont {Wan}, \citenamefont {Wang},\ and\
  \citenamefont {Yang}}]{shi_entanglement_2022}%
  \BibitemOpen
  \bibfield  {author} {\bibinfo {author} {\bibfnamefont {H.-L.}\ \bibnamefont
  {Shi}}, \bibinfo {author} {\bibfnamefont {S.}~\bibnamefont {Ding}}, \bibinfo
  {author} {\bibfnamefont {Q.-K.}\ \bibnamefont {Wan}}, \bibinfo {author}
  {\bibfnamefont {X.-H.}\ \bibnamefont {Wang}},\ and\ \bibinfo {author}
  {\bibfnamefont {W.-L.}\ \bibnamefont {Yang}},\ }\bibfield  {title} {\bibinfo
  {title} {Entanglement, {Coherence}, and {Extractable} {Work} in {Quantum}
  {Batteries}},\ }\href {https://doi.org/10.1103/PhysRevLett.129.130602}
  {\bibfield  {journal} {\bibinfo  {journal} {Phys. Rev. Lett.}\ }\textbf
  {\bibinfo {volume} {129}},\ \bibinfo {pages} {130602} (\bibinfo {year}
  {2022})}\BibitemShut {NoStop}%
\bibitem [{\citenamefont {Andolina}\ \emph
  {et~al.}(2019{\natexlab{a}})\citenamefont {Andolina}, \citenamefont {Keck},
  \citenamefont {Mari}, \citenamefont {Giovannetti},\ and\ \citenamefont
  {Polini}}]{andolina_quantum_2019}%
  \BibitemOpen
  \bibfield  {author} {\bibinfo {author} {\bibfnamefont {G.~M.}\ \bibnamefont
  {Andolina}}, \bibinfo {author} {\bibfnamefont {M.}~\bibnamefont {Keck}},
  \bibinfo {author} {\bibfnamefont {A.}~\bibnamefont {Mari}}, \bibinfo {author}
  {\bibfnamefont {V.}~\bibnamefont {Giovannetti}},\ and\ \bibinfo {author}
  {\bibfnamefont {M.}~\bibnamefont {Polini}},\ }\bibfield  {title} {\bibinfo
  {title} {Quantum versus classical many-body batteries},\ }\href
  {https://doi.org/10.1103/PhysRevB.99.205437} {\bibfield  {journal} {\bibinfo
  {journal} {Phys. Rev. B}\ }\textbf {\bibinfo {volume} {99}},\ \bibinfo
  {pages} {205437} (\bibinfo {year} {2019}{\natexlab{a}})}\BibitemShut
  {NoStop}%
\bibitem [{\citenamefont {Dou}\ \emph {et~al.}(2022)\citenamefont {Dou},
  \citenamefont {Lu}, \citenamefont {Wang},\ and\ \citenamefont
  {Sun}}]{dou_extended_2022}%
  \BibitemOpen
  \bibfield  {author} {\bibinfo {author} {\bibfnamefont {F.-Q.}\ \bibnamefont
  {Dou}}, \bibinfo {author} {\bibfnamefont {Y.-Q.}\ \bibnamefont {Lu}},
  \bibinfo {author} {\bibfnamefont {Y.-J.}\ \bibnamefont {Wang}},\ and\
  \bibinfo {author} {\bibfnamefont {J.-A.}\ \bibnamefont {Sun}},\ }\bibfield
  {title} {\bibinfo {title} {Extended dicke quantum battery with interatomic
  interactions and driving field},\ }\href
  {https://doi.org/10.1103/PhysRevB.105.115405} {\bibfield  {journal} {\bibinfo
   {journal} {Phys. Rev. B}\ }\textbf {\bibinfo {volume} {105}},\ \bibinfo
  {pages} {115405} (\bibinfo {year} {2022})}\BibitemShut {NoStop}%
\bibitem [{\citenamefont {Seidov}\ and\ \citenamefont
  {Mukhin}(2024)}]{seidov_quantum_2024}%
  \BibitemOpen
  \bibfield  {author} {\bibinfo {author} {\bibfnamefont {S.~S.}\ \bibnamefont
  {Seidov}}\ and\ \bibinfo {author} {\bibfnamefont {S.~I.}\ \bibnamefont
  {Mukhin}},\ }\bibfield  {title} {\bibinfo {title} {Quantum {Dicke} battery
  supercharging in the bound-luminosity state},\ }\href
  {https://doi.org/10.1103/PhysRevA.109.022210} {\bibfield  {journal} {\bibinfo
   {journal} {Phys. Rev. A}\ }\textbf {\bibinfo {volume} {109}},\ \bibinfo
  {pages} {022210} (\bibinfo {year} {2024})}\BibitemShut {NoStop}%
\bibitem [{\citenamefont {Li}\ and\ \citenamefont
  {Batchelor}(2021{\natexlab{a}})}]{PhysRevA.103.023719}%
  \BibitemOpen
  \bibfield  {author} {\bibinfo {author} {\bibfnamefont {Z.-M.}\ \bibnamefont
  {Li}}\ and\ \bibinfo {author} {\bibfnamefont {M.~T.}\ \bibnamefont
  {Batchelor}},\ }\bibfield  {title} {\bibinfo {title} {Hidden symmetry and
  tunneling dynamics in asymmetric quantum rabi models},\ }\href
  {https://doi.org/10.1103/PhysRevA.103.023719} {\bibfield  {journal} {\bibinfo
   {journal} {Phys. Rev. A}\ }\textbf {\bibinfo {volume} {103}},\ \bibinfo
  {pages} {023719} (\bibinfo {year} {2021}{\natexlab{a}})}\BibitemShut
  {NoStop}%
\bibitem [{\citenamefont {Li}\ \emph {et~al.}(2021{\natexlab{a}})\citenamefont
  {Li}, \citenamefont {Ferri},\ and\ \citenamefont
  {Batchelor}}]{PhysRevA.103.013711}%
  \BibitemOpen
  \bibfield  {author} {\bibinfo {author} {\bibfnamefont {Z.-M.}\ \bibnamefont
  {Li}}, \bibinfo {author} {\bibfnamefont {D.}~\bibnamefont {Ferri}},\ and\
  \bibinfo {author} {\bibfnamefont {M.~T.}\ \bibnamefont {Batchelor}},\
  }\bibfield  {title} {\bibinfo {title} {Nonorthogonal-qubit-state expansion
  for the asymmetric quantum rabi model},\ }\href
  {https://doi.org/10.1103/PhysRevA.103.013711} {\bibfield  {journal} {\bibinfo
   {journal} {Phys. Rev. A}\ }\textbf {\bibinfo {volume} {103}},\ \bibinfo
  {pages} {013711} (\bibinfo {year} {2021}{\natexlab{a}})}\BibitemShut
  {NoStop}%
\bibitem [{\citenamefont {Li}\ and\ \citenamefont
  {Batchelor}(2021{\natexlab{b}})}]{PhysRevA.104.033712}%
  \BibitemOpen
  \bibfield  {author} {\bibinfo {author} {\bibfnamefont {Z.-M.}\ \bibnamefont
  {Li}}\ and\ \bibinfo {author} {\bibfnamefont {M.~T.}\ \bibnamefont
  {Batchelor}},\ }\bibfield  {title} {\bibinfo {title} {Generalized adiabatic
  approximation to the quantum rabi model},\ }\href
  {https://doi.org/10.1103/PhysRevA.104.033712} {\bibfield  {journal} {\bibinfo
   {journal} {Phys. Rev. A}\ }\textbf {\bibinfo {volume} {104}},\ \bibinfo
  {pages} {033712} (\bibinfo {year} {2021}{\natexlab{b}})}\BibitemShut
  {NoStop}%
\bibitem [{\citenamefont {Li}\ and\ \citenamefont {Batchelor}(2015)}]{Li_2015}%
  \BibitemOpen
  \bibfield  {author} {\bibinfo {author} {\bibfnamefont {Z.-M.}\ \bibnamefont
  {Li}}\ and\ \bibinfo {author} {\bibfnamefont {M.~T.}\ \bibnamefont
  {Batchelor}},\ }\bibfield  {title} {\bibinfo {title} {Algebraic equations for
  the exceptional eigenspectrum of the generalized rabi model},\ }\href
  {https://doi.org/10.1088/1751-8113/48/45/454005} {\bibfield  {journal}
  {\bibinfo  {journal} {Journal of Physics A: Mathematical and Theoretical}\
  }\textbf {\bibinfo {volume} {48}},\ \bibinfo {pages} {454005} (\bibinfo
  {year} {2015})}\BibitemShut {NoStop}%
\bibitem [{\citenamefont {Xie}\ \emph {et~al.}(2017)\citenamefont {Xie},
  \citenamefont {Zhong}, \citenamefont {Batchelor},\ and\ \citenamefont
  {Lee}}]{Xie_2017}%
  \BibitemOpen
  \bibfield  {author} {\bibinfo {author} {\bibfnamefont {Q.}~\bibnamefont
  {Xie}}, \bibinfo {author} {\bibfnamefont {H.}~\bibnamefont {Zhong}}, \bibinfo
  {author} {\bibfnamefont {M.~T.}\ \bibnamefont {Batchelor}},\ and\ \bibinfo
  {author} {\bibfnamefont {C.}~\bibnamefont {Lee}},\ }\bibfield  {title}
  {\bibinfo {title} {The quantum rabi model: solution and dynamics},\ }\href
  {https://doi.org/10.1088/1751-8121/aa5a65} {\bibfield  {journal} {\bibinfo
  {journal} {Journal of Physics A: Mathematical and Theoretical}\ }\textbf
  {\bibinfo {volume} {50}},\ \bibinfo {pages} {113001} (\bibinfo {year}
  {2017})}\BibitemShut {NoStop}%
\bibitem [{\citenamefont {Li}\ \emph {et~al.}(2021{\natexlab{b}})\citenamefont
  {Li}, \citenamefont {Ferri}, \citenamefont {Tilbrook},\ and\ \citenamefont
  {Batchelor}}]{Li_2021}%
  \BibitemOpen
  \bibfield  {author} {\bibinfo {author} {\bibfnamefont {Z.-M.}\ \bibnamefont
  {Li}}, \bibinfo {author} {\bibfnamefont {D.}~\bibnamefont {Ferri}}, \bibinfo
  {author} {\bibfnamefont {D.}~\bibnamefont {Tilbrook}},\ and\ \bibinfo
  {author} {\bibfnamefont {M.~T.}\ \bibnamefont {Batchelor}},\ }\bibfield
  {title} {\bibinfo {title} {Generalized adiabatic approximation to the
  asymmetric quantum rabi model: conical intersections and geometric phases},\
  }\href {https://doi.org/10.1088/1751-8121/ac1fc1} {\bibfield  {journal}
  {\bibinfo  {journal} {Journal of Physics A: Mathematical and Theoretical}\
  }\textbf {\bibinfo {volume} {54}},\ \bibinfo {pages} {405201} (\bibinfo
  {year} {2021}{\natexlab{b}})}\BibitemShut {NoStop}%
\bibitem [{\citenamefont {Le}\ \emph {et~al.}(2018)\citenamefont {Le},
  \citenamefont {Levinsen}, \citenamefont {Modi}, \citenamefont {Parish},\ and\
  \citenamefont {Pollock}}]{Le_Spin_2018}%
  \BibitemOpen
  \bibfield  {author} {\bibinfo {author} {\bibfnamefont {T.~P.}\ \bibnamefont
  {Le}}, \bibinfo {author} {\bibfnamefont {J.}~\bibnamefont {Levinsen}},
  \bibinfo {author} {\bibfnamefont {K.}~\bibnamefont {Modi}}, \bibinfo {author}
  {\bibfnamefont {M.~M.}\ \bibnamefont {Parish}},\ and\ \bibinfo {author}
  {\bibfnamefont {F.~A.}\ \bibnamefont {Pollock}},\ }\bibfield  {title}
  {\bibinfo {title} {Spin-chain model of a many-body quantum battery},\ }\href
  {https://doi.org/10.1103/PhysRevA.97.022106} {\bibfield  {journal} {\bibinfo
  {journal} {Phys. Rev. A}\ }\textbf {\bibinfo {volume} {97}},\ \bibinfo
  {pages} {022106} (\bibinfo {year} {2018})}\BibitemShut {NoStop}%
\bibitem [{\citenamefont {Quach}\ \emph {et~al.}(2022)\citenamefont {Quach},
  \citenamefont {McGhee}, \citenamefont {Ganzer}, \citenamefont {Rouse},
  \citenamefont {Lovett}, \citenamefont {Gauger}, \citenamefont {Keeling},
  \citenamefont {Cerullo}, \citenamefont {Lidzey},\ and\ \citenamefont
  {Virgili}}]{James_Superabsorption_2022}%
  \BibitemOpen
  \bibfield  {author} {\bibinfo {author} {\bibfnamefont {J.~Q.}\ \bibnamefont
  {Quach}}, \bibinfo {author} {\bibfnamefont {K.~E.}\ \bibnamefont {McGhee}},
  \bibinfo {author} {\bibfnamefont {L.}~\bibnamefont {Ganzer}}, \bibinfo
  {author} {\bibfnamefont {D.~M.}\ \bibnamefont {Rouse}}, \bibinfo {author}
  {\bibfnamefont {B.~W.}\ \bibnamefont {Lovett}}, \bibinfo {author}
  {\bibfnamefont {E.~M.}\ \bibnamefont {Gauger}}, \bibinfo {author}
  {\bibfnamefont {J.}~\bibnamefont {Keeling}}, \bibinfo {author} {\bibfnamefont
  {G.}~\bibnamefont {Cerullo}}, \bibinfo {author} {\bibfnamefont {D.~G.}\
  \bibnamefont {Lidzey}},\ and\ \bibinfo {author} {\bibfnamefont
  {T.}~\bibnamefont {Virgili}},\ }\bibfield  {title} {\bibinfo {title}
  {Superabsorption in an organic microcavity: Toward a quantum battery},\
  }\href {https://doi.org/10.1126/sciadv.abk3160} {\bibfield  {journal}
  {\bibinfo  {journal} {Science Advances}\ }\textbf {\bibinfo {volume} {8}},\
  \bibinfo {pages} {eabk3160} (\bibinfo {year} {2022})}\BibitemShut {NoStop}%
\bibitem [{\citenamefont {Ferraro}\ \emph {et~al.}(2018)\citenamefont
  {Ferraro}, \citenamefont {Campisi}, \citenamefont {Andolina}, \citenamefont
  {Pellegrini},\ and\ \citenamefont {Polini}}]{Ferraro_High_2018}%
  \BibitemOpen
  \bibfield  {author} {\bibinfo {author} {\bibfnamefont {D.}~\bibnamefont
  {Ferraro}}, \bibinfo {author} {\bibfnamefont {M.}~\bibnamefont {Campisi}},
  \bibinfo {author} {\bibfnamefont {G.~M.}\ \bibnamefont {Andolina}}, \bibinfo
  {author} {\bibfnamefont {V.}~\bibnamefont {Pellegrini}},\ and\ \bibinfo
  {author} {\bibfnamefont {M.}~\bibnamefont {Polini}},\ }\bibfield  {title}
  {\bibinfo {title} {High-power collective charging of a solid-state quantum
  battery},\ }\href {https://doi.org/10.1103/PhysRevLett.120.117702} {\bibfield
   {journal} {\bibinfo  {journal} {Phys. Rev. Lett.}\ }\textbf {\bibinfo
  {volume} {120}},\ \bibinfo {pages} {117702} (\bibinfo {year}
  {2018})}\BibitemShut {NoStop}%
\bibitem [{\citenamefont {Andolina}\ \emph
  {et~al.}(2019{\natexlab{b}})\citenamefont {Andolina}, \citenamefont {Keck},
  \citenamefont {Mari}, \citenamefont {Campisi}, \citenamefont {Giovannetti},\
  and\ \citenamefont {Polini}}]{Andolina_Extractable_2019}%
  \BibitemOpen
  \bibfield  {author} {\bibinfo {author} {\bibfnamefont {G.~M.}\ \bibnamefont
  {Andolina}}, \bibinfo {author} {\bibfnamefont {M.}~\bibnamefont {Keck}},
  \bibinfo {author} {\bibfnamefont {A.}~\bibnamefont {Mari}}, \bibinfo {author}
  {\bibfnamefont {M.}~\bibnamefont {Campisi}}, \bibinfo {author} {\bibfnamefont
  {V.}~\bibnamefont {Giovannetti}},\ and\ \bibinfo {author} {\bibfnamefont
  {M.}~\bibnamefont {Polini}},\ }\bibfield  {title} {\bibinfo {title}
  {Extractable work, the role of correlations, and asymptotic freedom in
  quantum batteries},\ }\href {https://doi.org/10.1103/PhysRevLett.122.047702}
  {\bibfield  {journal} {\bibinfo  {journal} {Phys. Rev. Lett.}\ }\textbf
  {\bibinfo {volume} {122}},\ \bibinfo {pages} {047702} (\bibinfo {year}
  {2019}{\natexlab{b}})}\BibitemShut {NoStop}%
\bibitem [{\citenamefont {Juli\`a-Farr\'e}\ \emph {et~al.}(2020)\citenamefont
  {Juli\`a-Farr\'e}, \citenamefont {Salamon}, \citenamefont {Riera},
  \citenamefont {Bera},\ and\ \citenamefont {Lewenstein}}]{Juli_Bounds_2020}%
  \BibitemOpen
  \bibfield  {author} {\bibinfo {author} {\bibfnamefont {S.}~\bibnamefont
  {Juli\`a-Farr\'e}}, \bibinfo {author} {\bibfnamefont {T.}~\bibnamefont
  {Salamon}}, \bibinfo {author} {\bibfnamefont {A.}~\bibnamefont {Riera}},
  \bibinfo {author} {\bibfnamefont {M.~N.}\ \bibnamefont {Bera}},\ and\
  \bibinfo {author} {\bibfnamefont {M.}~\bibnamefont {Lewenstein}},\ }\bibfield
   {title} {\bibinfo {title} {Bounds on the capacity and power of quantum
  batteries},\ }\href {https://doi.org/10.1103/PhysRevResearch.2.023113}
  {\bibfield  {journal} {\bibinfo  {journal} {Phys. Rev. Res.}\ }\textbf
  {\bibinfo {volume} {2}},\ \bibinfo {pages} {023113} (\bibinfo {year}
  {2020})}\BibitemShut {NoStop}%
\bibitem [{\citenamefont {Caravelli}\ \emph {et~al.}(2020)\citenamefont
  {Caravelli}, \citenamefont {Coulter-De~Wit}, \citenamefont
  {Garc\'{\i}a-Pintos},\ and\ \citenamefont {Hamma}}]{Caravelli_Random_2020}%
  \BibitemOpen
  \bibfield  {author} {\bibinfo {author} {\bibfnamefont {F.}~\bibnamefont
  {Caravelli}}, \bibinfo {author} {\bibfnamefont {G.}~\bibnamefont
  {Coulter-De~Wit}}, \bibinfo {author} {\bibfnamefont {L.~P.}\ \bibnamefont
  {Garc\'{\i}a-Pintos}},\ and\ \bibinfo {author} {\bibfnamefont
  {A.}~\bibnamefont {Hamma}},\ }\bibfield  {title} {\bibinfo {title} {Random
  quantum batteries},\ }\href
  {https://doi.org/10.1103/PhysRevResearch.2.023095} {\bibfield  {journal}
  {\bibinfo  {journal} {Phys. Rev. Res.}\ }\textbf {\bibinfo {volume} {2}},\
  \bibinfo {pages} {023095} (\bibinfo {year} {2020})}\BibitemShut {NoStop}%
\bibitem [{\citenamefont {Crescente}\ \emph {et~al.}(2020)\citenamefont
  {Crescente}, \citenamefont {Carrega}, \citenamefont {Sassetti},\ and\
  \citenamefont {Ferraro}}]{crescente_ultrafast_2020}%
  \BibitemOpen
  \bibfield  {author} {\bibinfo {author} {\bibfnamefont {A.}~\bibnamefont
  {Crescente}}, \bibinfo {author} {\bibfnamefont {M.}~\bibnamefont {Carrega}},
  \bibinfo {author} {\bibfnamefont {M.}~\bibnamefont {Sassetti}},\ and\
  \bibinfo {author} {\bibfnamefont {D.}~\bibnamefont {Ferraro}},\ }\bibfield
  {title} {\bibinfo {title} {Ultrafast charging in a two-photon {Dicke} quantum
  battery},\ }\href {https://doi.org/10.1103/PhysRevB.102.245407} {\bibfield
  {journal} {\bibinfo  {journal} {Phys. Rev. B}\ }\textbf {\bibinfo {volume}
  {102}},\ \bibinfo {pages} {245407} (\bibinfo {year} {2020})}\BibitemShut
  {NoStop}%
\bibitem [{\citenamefont {Pokhrel}\ and\ \citenamefont
  {Gea-Banacloche}(2025)}]{PhysRevLett.134.130401}%
  \BibitemOpen
  \bibfield  {author} {\bibinfo {author} {\bibfnamefont {S.}~\bibnamefont
  {Pokhrel}}\ and\ \bibinfo {author} {\bibfnamefont {J.}~\bibnamefont
  {Gea-Banacloche}},\ }\bibfield  {title} {\bibinfo {title} {Large collective
  power enhancement in dissipative charging of a quantum battery},\ }\href
  {https://doi.org/10.1103/PhysRevLett.134.130401} {\bibfield  {journal}
  {\bibinfo  {journal} {Phys. Rev. Lett.}\ }\textbf {\bibinfo {volume} {134}},\
  \bibinfo {pages} {130401} (\bibinfo {year} {2025})}\BibitemShut {NoStop}%
\bibitem [{\citenamefont {Mayo}\ and\ \citenamefont
  {Roncaglia}(2022)}]{mayo_collective_2022}%
  \BibitemOpen
  \bibfield  {author} {\bibinfo {author} {\bibfnamefont {F.}~\bibnamefont
  {Mayo}}\ and\ \bibinfo {author} {\bibfnamefont {A.~J.}\ \bibnamefont
  {Roncaglia}},\ }\bibfield  {title} {\bibinfo {title} {Collective effects and
  quantum coherence in dissipative charging of quantum batteries},\ }\href
  {https://doi.org/10.1103/PhysRevA.105.062203} {\bibfield  {journal} {\bibinfo
   {journal} {Phys. Rev. A}\ }\textbf {\bibinfo {volume} {105}},\ \bibinfo
  {pages} {062203} (\bibinfo {year} {2022})}\BibitemShut {NoStop}%
\bibitem [{\citenamefont {Fusco}\ \emph {et~al.}(2016)\citenamefont {Fusco},
  \citenamefont {Paternostro},\ and\ \citenamefont
  {De~Chiara}}]{fusco_work_2016}%
  \BibitemOpen
  \bibfield  {author} {\bibinfo {author} {\bibfnamefont {L.}~\bibnamefont
  {Fusco}}, \bibinfo {author} {\bibfnamefont {M.}~\bibnamefont {Paternostro}},\
  and\ \bibinfo {author} {\bibfnamefont {G.}~\bibnamefont {De~Chiara}},\
  }\bibfield  {title} {\bibinfo {title} {Work extraction and energy storage in
  the dicke model},\ }\href {https://doi.org/10.1103/PhysRevE.94.052122}
  {\bibfield  {journal} {\bibinfo  {journal} {Phys. Rev. E}\ }\textbf {\bibinfo
  {volume} {94}},\ \bibinfo {pages} {052122} (\bibinfo {year}
  {2016})}\BibitemShut {NoStop}%
\bibitem [{\citenamefont {Lu}\ \emph {et~al.}(2021)\citenamefont {Lu},
  \citenamefont {Li}, \citenamefont {Mangazeev},\ and\ \citenamefont
  {Batchelor}}]{Lu_2021}%
  \BibitemOpen
  \bibfield  {author} {\bibinfo {author} {\bibfnamefont {X.}~\bibnamefont
  {Lu}}, \bibinfo {author} {\bibfnamefont {Z.-M.}\ \bibnamefont {Li}}, \bibinfo
  {author} {\bibfnamefont {V.~V.}\ \bibnamefont {Mangazeev}},\ and\ \bibinfo
  {author} {\bibfnamefont {M.~T.}\ \bibnamefont {Batchelor}},\ }\bibfield
  {title} {\bibinfo {title} {Hidden symmetry in the biased dicke model},\
  }\href {https://doi.org/10.1088/1751-8121/ac0f16} {\bibfield  {journal}
  {\bibinfo  {journal} {Journal of Physics A: Mathematical and Theoretical}\
  }\textbf {\bibinfo {volume} {54}},\ \bibinfo {pages} {325202} (\bibinfo
  {year} {2021})}\BibitemShut {NoStop}%
\bibitem [{\citenamefont {Gyhm}\ \emph {et~al.}(2022)\citenamefont {Gyhm},
  \citenamefont {\ifmmode~\check{S}\else \v{S}\fi{}afr\'anek},\ and\
  \citenamefont {Rosa}}]{Gyhm_Quantum_2022}%
  \BibitemOpen
  \bibfield  {author} {\bibinfo {author} {\bibfnamefont {J.-Y.}\ \bibnamefont
  {Gyhm}}, \bibinfo {author} {\bibfnamefont {D.}~\bibnamefont
  {\ifmmode~\check{S}\else \v{S}\fi{}afr\'anek}},\ and\ \bibinfo {author}
  {\bibfnamefont {D.}~\bibnamefont {Rosa}},\ }\bibfield  {title} {\bibinfo
  {title} {Quantum charging advantage cannot be extensive without global
  operations},\ }\href {https://doi.org/10.1103/PhysRevLett.128.140501}
  {\bibfield  {journal} {\bibinfo  {journal} {Phys. Rev. Lett.}\ }\textbf
  {\bibinfo {volume} {128}},\ \bibinfo {pages} {140501} (\bibinfo {year}
  {2022})}\BibitemShut {NoStop}%
\bibitem [{\citenamefont {Kamin}\ \emph {et~al.}(2020)\citenamefont {Kamin},
  \citenamefont {Tabesh}, \citenamefont {Salimi}, \citenamefont {Kheirandish},\
  and\ \citenamefont {Santos}}]{Kamin_Non_2020}%
  \BibitemOpen
  \bibfield  {author} {\bibinfo {author} {\bibfnamefont {F.~H.}\ \bibnamefont
  {Kamin}}, \bibinfo {author} {\bibfnamefont {F.~T.}\ \bibnamefont {Tabesh}},
  \bibinfo {author} {\bibfnamefont {S.}~\bibnamefont {Salimi}}, \bibinfo
  {author} {\bibfnamefont {F.}~\bibnamefont {Kheirandish}},\ and\ \bibinfo
  {author} {\bibfnamefont {A.~C.}\ \bibnamefont {Santos}},\ }\bibfield  {title}
  {\bibinfo {title} {Non-markovian effects on charging and self-discharging
  process of quantum batteries},\ }\href
  {https://doi.org/10.1088/1367-2630/ab9ee2} {\bibfield  {journal} {\bibinfo
  {journal} {New J. Phys.}\ }\textbf {\bibinfo {volume} {22}},\ \bibinfo
  {pages} {083007} (\bibinfo {year} {2020})}\BibitemShut {NoStop}%
\bibitem [{\citenamefont {Pirmoradian}\ and\ \citenamefont
  {M\o{}lmer}(2019)}]{Pirmoradian_Aging_2019}%
  \BibitemOpen
  \bibfield  {author} {\bibinfo {author} {\bibfnamefont {F.}~\bibnamefont
  {Pirmoradian}}\ and\ \bibinfo {author} {\bibfnamefont {K.}~\bibnamefont
  {M\o{}lmer}},\ }\bibfield  {title} {\bibinfo {title} {Aging of a quantum
  battery},\ }\href {https://doi.org/10.1103/PhysRevA.100.043833} {\bibfield
  {journal} {\bibinfo  {journal} {Phys. Rev. A}\ }\textbf {\bibinfo {volume}
  {100}},\ \bibinfo {pages} {043833} (\bibinfo {year} {2019})}\BibitemShut
  {NoStop}%
\bibitem [{\citenamefont {Quach}\ and\ \citenamefont
  {Munro}(2020)}]{Quach_Using_2020}%
  \BibitemOpen
  \bibfield  {author} {\bibinfo {author} {\bibfnamefont {J.~Q.}\ \bibnamefont
  {Quach}}\ and\ \bibinfo {author} {\bibfnamefont {W.~J.}\ \bibnamefont
  {Munro}},\ }\bibfield  {title} {\bibinfo {title} {Using dark states to charge
  and stabilize open quantum batteries},\ }\href
  {https://doi.org/10.1103/PhysRevApplied.14.024092} {\bibfield  {journal}
  {\bibinfo  {journal} {Phys. Rev. Appl.}\ }\textbf {\bibinfo {volume} {14}},\
  \bibinfo {pages} {024092} (\bibinfo {year} {2020})}\BibitemShut {NoStop}%
\bibitem [{\citenamefont {Zhang}\ \emph {et~al.}(2019)\citenamefont {Zhang},
  \citenamefont {Yang}, \citenamefont {Fu},\ and\ \citenamefont
  {Wang}}]{Zhang_Powerful_2019}%
  \BibitemOpen
  \bibfield  {author} {\bibinfo {author} {\bibfnamefont {Y.-Y.}\ \bibnamefont
  {Zhang}}, \bibinfo {author} {\bibfnamefont {T.-R.}\ \bibnamefont {Yang}},
  \bibinfo {author} {\bibfnamefont {L.}~\bibnamefont {Fu}},\ and\ \bibinfo
  {author} {\bibfnamefont {X.}~\bibnamefont {Wang}},\ }\bibfield  {title}
  {\bibinfo {title} {Powerful harmonic charging in a quantum battery},\ }\href
  {https://doi.org/10.1103/PhysRevE.99.052106} {\bibfield  {journal} {\bibinfo
  {journal} {Phys. Rev. E}\ }\textbf {\bibinfo {volume} {99}},\ \bibinfo
  {pages} {052106} (\bibinfo {year} {2019})}\BibitemShut {NoStop}%
\bibitem [{\citenamefont {Andolina}\ \emph {et~al.}(2018)\citenamefont
  {Andolina}, \citenamefont {Farina}, \citenamefont {Mari}, \citenamefont
  {Pellegrini}, \citenamefont {Giovannetti},\ and\ \citenamefont
  {Polini}}]{PhysRevB.98.205423}%
  \BibitemOpen
  \bibfield  {author} {\bibinfo {author} {\bibfnamefont {G.~M.}\ \bibnamefont
  {Andolina}}, \bibinfo {author} {\bibfnamefont {D.}~\bibnamefont {Farina}},
  \bibinfo {author} {\bibfnamefont {A.}~\bibnamefont {Mari}}, \bibinfo {author}
  {\bibfnamefont {V.}~\bibnamefont {Pellegrini}}, \bibinfo {author}
  {\bibfnamefont {V.}~\bibnamefont {Giovannetti}},\ and\ \bibinfo {author}
  {\bibfnamefont {M.}~\bibnamefont {Polini}},\ }\bibfield  {title} {\bibinfo
  {title} {Charger-mediated energy transfer in exactly solvable models for
  quantum batteries},\ }\href {https://doi.org/10.1103/PhysRevB.98.205423}
  {\bibfield  {journal} {\bibinfo  {journal} {Phys. Rev. B}\ }\textbf {\bibinfo
  {volume} {98}},\ \bibinfo {pages} {205423} (\bibinfo {year}
  {2018})}\BibitemShut {NoStop}%
\bibitem [{\citenamefont {Konar}\ \emph {et~al.}(2024)\citenamefont {Konar},
  \citenamefont {Lakkaraju},\ and\ \citenamefont
  {Sen~(De)}}]{PhysRevA.109.042207}%
  \BibitemOpen
  \bibfield  {author} {\bibinfo {author} {\bibfnamefont {T.~K.}\ \bibnamefont
  {Konar}}, \bibinfo {author} {\bibfnamefont {L.~G.~C.}\ \bibnamefont
  {Lakkaraju}},\ and\ \bibinfo {author} {\bibfnamefont {A.}~\bibnamefont
  {Sen~(De)}},\ }\bibfield  {title} {\bibinfo {title} {Quantum battery with
  non-hermitian charging},\ }\href
  {https://doi.org/10.1103/PhysRevA.109.042207} {\bibfield  {journal} {\bibinfo
   {journal} {Phys. Rev. A}\ }\textbf {\bibinfo {volume} {109}},\ \bibinfo
  {pages} {042207} (\bibinfo {year} {2024})}\BibitemShut {NoStop}%
\bibitem [{\citenamefont {Lin}\ \emph {et~al.}(2026)\citenamefont {Lin},
  \citenamefont {Ye}, \citenamefont {Li}, \citenamefont {Su},\ and\
  \citenamefont {Wu}}]{lin2026enhanced}%
  \BibitemOpen
  \bibfield  {author} {\bibinfo {author} {\bibfnamefont {Q.-Y.}\ \bibnamefont
  {Lin}}, \bibinfo {author} {\bibfnamefont {G.-Z.}\ \bibnamefont {Ye}},
  \bibinfo {author} {\bibfnamefont {C.}~\bibnamefont {Li}}, \bibinfo {author}
  {\bibfnamefont {W.-J.}\ \bibnamefont {Su}},\ and\ \bibinfo {author}
  {\bibfnamefont {H.-Z.}\ \bibnamefont {Wu}},\ }\bibfield  {title} {\bibinfo
  {title} {Enhanced charging power in nonreciprocal quantum battery by
  reservoir engineering},\ }\href {https://doi.org/10.1088/1402-4896/ae31bc}
  {\bibfield  {journal} {\bibinfo  {journal} {Physica Scripta}\ }\textbf
  {\bibinfo {volume} {101}},\ \bibinfo {pages} {025104} (\bibinfo {year}
  {2026})}\BibitemShut {NoStop}%
\bibitem [{\citenamefont {Yang}\ and\ \citenamefont {Dou}(2025)}]{8xsm-5mb6}%
  \BibitemOpen
  \bibfield  {author} {\bibinfo {author} {\bibfnamefont {F.-M.}\ \bibnamefont
  {Yang}}\ and\ \bibinfo {author} {\bibfnamefont {F.-Q.}\ \bibnamefont {Dou}},\
  }\bibfield  {title} {\bibinfo {title} {Wireless energy transfer in a
  non-hermitian quantum battery},\ }\href {https://doi.org/10.1103/8xsm-5mb6}
  {\bibfield  {journal} {\bibinfo  {journal} {Phys. Rev. A}\ }\textbf {\bibinfo
  {volume} {112}},\ \bibinfo {pages} {042205} (\bibinfo {year}
  {2025})}\BibitemShut {NoStop}%
\bibitem [{\citenamefont {Zhou}\ \emph {et~al.}(2026)\citenamefont {Zhou},
  \citenamefont {Xiao}, \citenamefont {Xu}, \citenamefont {Gao}, \citenamefont
  {Li}, \citenamefont {Zhou}, \citenamefont {Li},\ and\ \citenamefont
  {Shu}}]{4klp-kw27}%
  \BibitemOpen
  \bibfield  {author} {\bibinfo {author} {\bibfnamefont {A.-L.}\ \bibnamefont
  {Zhou}}, \bibinfo {author} {\bibfnamefont {Y.-W.}\ \bibnamefont {Xiao}},
  \bibinfo {author} {\bibfnamefont {N.}~\bibnamefont {Xu}}, \bibinfo {author}
  {\bibfnamefont {L.-L.}\ \bibnamefont {Gao}}, \bibinfo {author} {\bibfnamefont
  {L.-J.}\ \bibnamefont {Li}}, \bibinfo {author} {\bibfnamefont
  {H.}~\bibnamefont {Zhou}}, \bibinfo {author} {\bibfnamefont {Z.-M.}\
  \bibnamefont {Li}},\ and\ \bibinfo {author} {\bibfnamefont {C.-C.}\
  \bibnamefont {Shu}},\ }\bibfield  {title} {\bibinfo {title} {Topological
  enhancement of a $\mathcal{PT}$-symmetric su-schrieffer-heeger quantum
  battery},\ }\href {https://doi.org/10.1103/4klp-kw27} {\bibfield  {journal}
  {\bibinfo  {journal} {Phys. Rev. A}\ }\textbf {\bibinfo {volume} {113}},\
  \bibinfo {pages} {042213} (\bibinfo {year} {2026})}\BibitemShut {NoStop}%
\bibitem [{\citenamefont {Hovhannisyan}\ \emph {et~al.}(2020)\citenamefont
  {Hovhannisyan}, \citenamefont {Barra},\ and\ \citenamefont
  {Imparato}}]{hovhannisyan_charging_2020}%
  \BibitemOpen
  \bibfield  {author} {\bibinfo {author} {\bibfnamefont {K.~V.}\ \bibnamefont
  {Hovhannisyan}}, \bibinfo {author} {\bibfnamefont {F.}~\bibnamefont
  {Barra}},\ and\ \bibinfo {author} {\bibfnamefont {A.}~\bibnamefont
  {Imparato}},\ }\bibfield  {title} {\bibinfo {title} {Charging assisted by
  thermalization},\ }\href {https://doi.org/10.1103/PhysRevResearch.2.033413}
  {\bibfield  {journal} {\bibinfo  {journal} {Phys. Rev. Research}\ }\textbf
  {\bibinfo {volume} {2}},\ \bibinfo {pages} {033413} (\bibinfo {year}
  {2020})}\BibitemShut {NoStop}%
\bibitem [{\citenamefont {Ashida}\ \emph {et~al.}(2020)\citenamefont {Ashida},
  \citenamefont {Gong},\ and\ \citenamefont {Ueda}}]{Ashida02072020}%
  \BibitemOpen
  \bibfield  {author} {\bibinfo {author} {\bibfnamefont {Y.}~\bibnamefont
  {Ashida}}, \bibinfo {author} {\bibfnamefont {Z.}~\bibnamefont {Gong}},\ and\
  \bibinfo {author} {\bibfnamefont {M.}~\bibnamefont {Ueda}},\ }\bibfield
  {title} {\bibinfo {title} {Non-hermitian physics},\ }\href
  {https://doi.org/10.1080/00018732.2021.1876991} {\bibfield  {journal}
  {\bibinfo  {journal} {Advances in Physics}\ }\textbf {\bibinfo {volume}
  {69}},\ \bibinfo {pages} {249} (\bibinfo {year} {2020})}\BibitemShut
  {NoStop}%
\bibitem [{\citenamefont {Roccati}\ \emph {et~al.}(2022)\citenamefont
  {Roccati}, \citenamefont {Palma}, \citenamefont {Ciccarello},\ and\
  \citenamefont {Bagarello}}]{RoccatiNon-Hermitian2022}%
  \BibitemOpen
  \bibfield  {author} {\bibinfo {author} {\bibfnamefont {F.}~\bibnamefont
  {Roccati}}, \bibinfo {author} {\bibfnamefont {G.~M.}\ \bibnamefont {Palma}},
  \bibinfo {author} {\bibfnamefont {F.}~\bibnamefont {Ciccarello}},\ and\
  \bibinfo {author} {\bibfnamefont {F.}~\bibnamefont {Bagarello}},\ }\bibfield
  {title} {\bibinfo {title} {Non-hermitian physics and master equations},\
  }\href {https://doi.org/10.1142/S1230161222500044} {\bibfield  {journal}
  {\bibinfo  {journal} {Open Systems \& Information Dynamics}\ }\textbf
  {\bibinfo {volume} {29}},\ \bibinfo {pages} {2250004} (\bibinfo {year}
  {2022})}\BibitemShut {NoStop}%
\bibitem [{\citenamefont {Li}\ \emph {et~al.}(2024)\citenamefont {Li},
  \citenamefont {Shi}, \citenamefont {Fan}, \citenamefont {Li},\ and\
  \citenamefont {Shu}}]{PhysRevA.110.023708}%
  \BibitemOpen
  \bibfield  {author} {\bibinfo {author} {\bibfnamefont {H.}~\bibnamefont
  {Li}}, \bibinfo {author} {\bibfnamefont {J.-K.}\ \bibnamefont {Shi}},
  \bibinfo {author} {\bibfnamefont {L.-B.}\ \bibnamefont {Fan}}, \bibinfo
  {author} {\bibfnamefont {Z.-M.}\ \bibnamefont {Li}},\ and\ \bibinfo {author}
  {\bibfnamefont {C.-C.}\ \bibnamefont {Shu}},\ }\bibfield  {title} {\bibinfo
  {title} {Anisotropic rabi model with two-photon relaxation},\ }\href
  {https://doi.org/10.1103/PhysRevA.110.023708} {\bibfield  {journal} {\bibinfo
   {journal} {Phys. Rev. A}\ }\textbf {\bibinfo {volume} {110}},\ \bibinfo
  {pages} {023708} (\bibinfo {year} {2024})}\BibitemShut {NoStop}%
\bibitem [{\citenamefont {Allahverdyan}\ \emph {et~al.}(2004)\citenamefont
  {Allahverdyan}, \citenamefont {Balian},\ and\ \citenamefont
  {Nieuwenhuizen}}]{allahverdyan_maximal_2004}%
  \BibitemOpen
  \bibfield  {author} {\bibinfo {author} {\bibfnamefont {A.~E.}\ \bibnamefont
  {Allahverdyan}}, \bibinfo {author} {\bibfnamefont {R.}~\bibnamefont
  {Balian}},\ and\ \bibinfo {author} {\bibfnamefont {T.~M.}\ \bibnamefont
  {Nieuwenhuizen}},\ }\bibfield  {title} {\bibinfo {title} {Maximal work
  extraction from finite quantum systems},\ }\href
  {https://doi.org/10.1209/epl/i2004-10101-2} {\bibfield  {journal} {\bibinfo
  {journal} {Europhys. Lett.}\ }\textbf {\bibinfo {volume} {67}},\ \bibinfo
  {pages} {565} (\bibinfo {year} {2004})}\BibitemShut {NoStop}%
\bibitem [{\citenamefont {Yang}\ \emph {et~al.}(2023)\citenamefont {Yang},
  \citenamefont {Yang}, \citenamefont {Alimuddin}, \citenamefont {Salvia},
  \citenamefont {Fei}, \citenamefont {Zhao}, \citenamefont {Nimmrichter},\ and\
  \citenamefont {Luo}}]{Yang_Battery_2023}%
  \BibitemOpen
  \bibfield  {author} {\bibinfo {author} {\bibfnamefont {X.}~\bibnamefont
  {Yang}}, \bibinfo {author} {\bibfnamefont {Y.-H.}\ \bibnamefont {Yang}},
  \bibinfo {author} {\bibfnamefont {M.}~\bibnamefont {Alimuddin}}, \bibinfo
  {author} {\bibfnamefont {R.}~\bibnamefont {Salvia}}, \bibinfo {author}
  {\bibfnamefont {S.-M.}\ \bibnamefont {Fei}}, \bibinfo {author} {\bibfnamefont
  {L.-M.}\ \bibnamefont {Zhao}}, \bibinfo {author} {\bibfnamefont
  {S.}~\bibnamefont {Nimmrichter}},\ and\ \bibinfo {author} {\bibfnamefont
  {M.-X.}\ \bibnamefont {Luo}},\ }\bibfield  {title} {\bibinfo {title} {Battery
  capacity of energy-storing quantum systems},\ }\href
  {https://doi.org/10.1103/PhysRevLett.131.030402} {\bibfield  {journal}
  {\bibinfo  {journal} {Phys. Rev. Lett.}\ }\textbf {\bibinfo {volume} {131}},\
  \bibinfo {pages} {030402} (\bibinfo {year} {2023})}\BibitemShut {NoStop}%
\bibitem [{\citenamefont {Naghiloo}\ \emph {et~al.}(2019)\citenamefont
  {Naghiloo}, \citenamefont {Abbasi}, \citenamefont {Joglekar},\ and\
  \citenamefont {Murch}}]{naghilooQuantum2019a}%
  \BibitemOpen
  \bibfield  {author} {\bibinfo {author} {\bibfnamefont {M.}~\bibnamefont
  {Naghiloo}}, \bibinfo {author} {\bibfnamefont {M.}~\bibnamefont {Abbasi}},
  \bibinfo {author} {\bibfnamefont {Y.~N.}\ \bibnamefont {Joglekar}},\ and\
  \bibinfo {author} {\bibfnamefont {K.~W.}\ \bibnamefont {Murch}},\ }\bibfield
  {title} {\bibinfo {title} {Quantum state tomography across the exceptional
  point in a single dissipative qubit},\ }\href
  {https://doi.org/10.1038/s41567-019-0652-z} {\bibfield  {journal} {\bibinfo
  {journal} {Nature Physics}\ }\textbf {\bibinfo {volume} {15}},\ \bibinfo
  {pages} {1232} (\bibinfo {year} {2019})}\BibitemShut {NoStop}%
\bibitem [{\citenamefont {Li}\ \emph {et~al.}(2019)\citenamefont {Li},
  \citenamefont {Harter}, \citenamefont {Liu}, \citenamefont {de~Melo},
  \citenamefont {Joglekar},\ and\ \citenamefont {Luo}}]{liObservation2019a}%
  \BibitemOpen
  \bibfield  {author} {\bibinfo {author} {\bibfnamefont {J.}~\bibnamefont
  {Li}}, \bibinfo {author} {\bibfnamefont {A.~K.}\ \bibnamefont {Harter}},
  \bibinfo {author} {\bibfnamefont {J.}~\bibnamefont {Liu}}, \bibinfo {author}
  {\bibfnamefont {L.}~\bibnamefont {de~Melo}}, \bibinfo {author} {\bibfnamefont
  {Y.~N.}\ \bibnamefont {Joglekar}},\ and\ \bibinfo {author} {\bibfnamefont
  {L.}~\bibnamefont {Luo}},\ }\bibfield  {title} {\bibinfo {title} {Observation
  of parity-time symmetry breaking transitions in a dissipative floquet system
  of ultracold atoms},\ }\href {https://doi.org/10.1038/s41467-019-08596-1}
  {\bibfield  {journal} {\bibinfo  {journal} {Nature Communications}\ }\textbf
  {\bibinfo {volume} {10}},\ \bibinfo {pages} {855} (\bibinfo {year}
  {2019})}\BibitemShut {NoStop}%
\bibitem [{\citenamefont {Lu}\ \emph {et~al.}(2023)\citenamefont {Lu},
  \citenamefont {Li}, \citenamefont {Shi}, \citenamefont {Fan}, \citenamefont
  {Mangazeev}, \citenamefont {Li},\ and\ \citenamefont
  {Batchelor}}]{PhysRevA.108.053712}%
  \BibitemOpen
  \bibfield  {author} {\bibinfo {author} {\bibfnamefont {X.}~\bibnamefont
  {Lu}}, \bibinfo {author} {\bibfnamefont {H.}~\bibnamefont {Li}}, \bibinfo
  {author} {\bibfnamefont {J.-K.}\ \bibnamefont {Shi}}, \bibinfo {author}
  {\bibfnamefont {L.-B.}\ \bibnamefont {Fan}}, \bibinfo {author} {\bibfnamefont
  {V.}~\bibnamefont {Mangazeev}}, \bibinfo {author} {\bibfnamefont {Z.-M.}\
  \bibnamefont {Li}},\ and\ \bibinfo {author} {\bibfnamefont {M.~T.}\
  \bibnamefont {Batchelor}},\ }\bibfield  {title} {\bibinfo {title}
  {$\mathcal{PT}$-symmetric quantum rabi model},\ }\href
  {https://doi.org/10.1103/PhysRevA.108.053712} {\bibfield  {journal} {\bibinfo
   {journal} {Phys. Rev. A}\ }\textbf {\bibinfo {volume} {108}},\ \bibinfo
  {pages} {053712} (\bibinfo {year} {2023})}\BibitemShut {NoStop}%
\bibitem [{\citenamefont {Baumann}\ \emph {et~al.}(2010)\citenamefont
  {Baumann}, \citenamefont {Guerlin}, \citenamefont {Brennecke},\ and\
  \citenamefont {Esslinger}}]{baumannDicke2010}%
  \BibitemOpen
  \bibfield  {author} {\bibinfo {author} {\bibfnamefont {K.}~\bibnamefont
  {Baumann}}, \bibinfo {author} {\bibfnamefont {C.}~\bibnamefont {Guerlin}},
  \bibinfo {author} {\bibfnamefont {F.}~\bibnamefont {Brennecke}},\ and\
  \bibinfo {author} {\bibfnamefont {T.}~\bibnamefont {Esslinger}},\ }\bibfield
  {title} {\bibinfo {title} {Dicke quantum phase transition with a superfluid
  gas in an optical cavity},\ }\href {https://doi.org/10.1038/nature09009}
  {\bibfield  {journal} {\bibinfo  {journal} {Nature}\ }\textbf {\bibinfo
  {volume} {464}},\ \bibinfo {pages} {1301} (\bibinfo {year}
  {2010})}\BibitemShut {NoStop}%
\bibitem [{\citenamefont {Hu}\ \emph {et~al.}(2022)\citenamefont {Hu},
  \citenamefont {Qiu}, \citenamefont {Souza}, \citenamefont {Yuan},
  \citenamefont {Zhou}, \citenamefont {Zhang}, \citenamefont {Chu},
  \citenamefont {Pan}, \citenamefont {Hu}, \citenamefont {Li}, \citenamefont
  {Xu}, \citenamefont {Zhong}, \citenamefont {Liu}, \citenamefont {Yan},
  \citenamefont {Tan}, \citenamefont {Bachelard}, \citenamefont {Villas-Boas},
  \citenamefont {Santos},\ and\ \citenamefont {Yu}}]{Hu_2022}%
  \BibitemOpen
  \bibfield  {author} {\bibinfo {author} {\bibfnamefont {C.-K.}\ \bibnamefont
  {Hu}}, \bibinfo {author} {\bibfnamefont {J.}~\bibnamefont {Qiu}}, \bibinfo
  {author} {\bibfnamefont {P.~J.~P.}\ \bibnamefont {Souza}}, \bibinfo {author}
  {\bibfnamefont {J.}~\bibnamefont {Yuan}}, \bibinfo {author} {\bibfnamefont
  {Y.}~\bibnamefont {Zhou}}, \bibinfo {author} {\bibfnamefont {L.}~\bibnamefont
  {Zhang}}, \bibinfo {author} {\bibfnamefont {J.}~\bibnamefont {Chu}}, \bibinfo
  {author} {\bibfnamefont {X.}~\bibnamefont {Pan}}, \bibinfo {author}
  {\bibfnamefont {L.}~\bibnamefont {Hu}}, \bibinfo {author} {\bibfnamefont
  {J.}~\bibnamefont {Li}}, \bibinfo {author} {\bibfnamefont {Y.}~\bibnamefont
  {Xu}}, \bibinfo {author} {\bibfnamefont {Y.}~\bibnamefont {Zhong}}, \bibinfo
  {author} {\bibfnamefont {S.}~\bibnamefont {Liu}}, \bibinfo {author}
  {\bibfnamefont {F.}~\bibnamefont {Yan}}, \bibinfo {author} {\bibfnamefont
  {D.}~\bibnamefont {Tan}}, \bibinfo {author} {\bibfnamefont {R.}~\bibnamefont
  {Bachelard}}, \bibinfo {author} {\bibfnamefont {C.~J.}\ \bibnamefont
  {Villas-Boas}}, \bibinfo {author} {\bibfnamefont {A.~C.}\ \bibnamefont
  {Santos}},\ and\ \bibinfo {author} {\bibfnamefont {D.}~\bibnamefont {Yu}},\
  }\bibfield  {title} {\bibinfo {title} {Optimal charging of a superconducting
  quantum battery},\ }\href {https://doi.org/10.1088/2058-9565/ac8444}
  {\bibfield  {journal} {\bibinfo  {journal} {Quantum Science and Technology}\
  }\textbf {\bibinfo {volume} {7}},\ \bibinfo {pages} {045018} (\bibinfo {year}
  {2022})}\BibitemShut {NoStop}%
\bibitem [{\citenamefont {Wang}\ \emph {et~al.}(2021)\citenamefont {Wang},
  \citenamefont {Zhou}, \citenamefont {Zhang}, \citenamefont {Zhang},
  \citenamefont {Zhang}, \citenamefont {Xie}, \citenamefont {Wu}, \citenamefont
  {Chen}, \citenamefont {Ou}, \citenamefont {Wu}, \citenamefont {Jing},\ and\
  \citenamefont {Chen}}]{PhysRevA.103.L020201}%
  \BibitemOpen
  \bibfield  {author} {\bibinfo {author} {\bibfnamefont {W.-C.}\ \bibnamefont
  {Wang}}, \bibinfo {author} {\bibfnamefont {Y.-L.}\ \bibnamefont {Zhou}},
  \bibinfo {author} {\bibfnamefont {H.-L.}\ \bibnamefont {Zhang}}, \bibinfo
  {author} {\bibfnamefont {J.}~\bibnamefont {Zhang}}, \bibinfo {author}
  {\bibfnamefont {M.-C.}\ \bibnamefont {Zhang}}, \bibinfo {author}
  {\bibfnamefont {Y.}~\bibnamefont {Xie}}, \bibinfo {author} {\bibfnamefont
  {C.-W.}\ \bibnamefont {Wu}}, \bibinfo {author} {\bibfnamefont
  {T.}~\bibnamefont {Chen}}, \bibinfo {author} {\bibfnamefont {B.-Q.}\
  \bibnamefont {Ou}}, \bibinfo {author} {\bibfnamefont {W.}~\bibnamefont {Wu}},
  \bibinfo {author} {\bibfnamefont {H.}~\bibnamefont {Jing}},\ and\ \bibinfo
  {author} {\bibfnamefont {P.-X.}\ \bibnamefont {Chen}},\ }\bibfield  {title}
  {\bibinfo {title} {Observation of $\mathcal{PT}$-symmetric quantum coherence
  in a single-ion system},\ }\href
  {https://doi.org/10.1103/PhysRevA.103.L020201} {\bibfield  {journal}
  {\bibinfo  {journal} {Phys. Rev. A}\ }\textbf {\bibinfo {volume} {103}},\
  \bibinfo {pages} {L020201} (\bibinfo {year} {2021})}\BibitemShut {NoStop}%
\bibitem [{\citenamefont {Huber}\ \emph {et~al.}(2020)\citenamefont {Huber},
  \citenamefont {Kirton}, \citenamefont {Rotter},\ and\ \citenamefont
  {Rabl}}]{huber_emergence_2020}%
  \BibitemOpen
  \bibfield  {author} {\bibinfo {author} {\bibfnamefont {J.}~\bibnamefont
  {Huber}}, \bibinfo {author} {\bibfnamefont {P.}~\bibnamefont {Kirton}},
  \bibinfo {author} {\bibfnamefont {S.}~\bibnamefont {Rotter}},\ and\ \bibinfo
  {author} {\bibfnamefont {P.}~\bibnamefont {Rabl}},\ }\bibfield  {title}
  {\bibinfo {title} {Emergence of {PT}-symmetry breaking in open quantum
  systems},\ }\href {https://doi.org/10.21468/SciPostPhys.9.4.052} {\bibfield
  {journal} {\bibinfo  {journal} {SciPost Phys.}\ }\textbf {\bibinfo {volume}
  {9}},\ \bibinfo {pages} {052} (\bibinfo {year} {2020})}\BibitemShut {NoStop}%
\bibitem [{\citenamefont {Dast}\ \emph {et~al.}(2014)\citenamefont {Dast},
  \citenamefont {Haag}, \citenamefont {Cartarius},\ and\ \citenamefont
  {Wunner}}]{dast_quantum_2014}%
  \BibitemOpen
  \bibfield  {author} {\bibinfo {author} {\bibfnamefont {D.}~\bibnamefont
  {Dast}}, \bibinfo {author} {\bibfnamefont {D.}~\bibnamefont {Haag}}, \bibinfo
  {author} {\bibfnamefont {H.}~\bibnamefont {Cartarius}},\ and\ \bibinfo
  {author} {\bibfnamefont {G.}~\bibnamefont {Wunner}},\ }\bibfield  {title}
  {\bibinfo {title} {Quantum master equation with balanced gain and loss},\
  }\href {https://doi.org/10.1103/PhysRevA.90.052120} {\bibfield  {journal}
  {\bibinfo  {journal} {Phys. Rev. A}\ }\textbf {\bibinfo {volume} {90}},\
  \bibinfo {pages} {052120} (\bibinfo {year} {2014})}\BibitemShut {NoStop}%
\bibitem [{\citenamefont {Manzano}(2020)}]{ManzanoAIP}%
  \BibitemOpen
  \bibfield  {author} {\bibinfo {author} {\bibfnamefont {D.}~\bibnamefont
  {Manzano}},\ }\bibfield  {title} {\bibinfo {title} {A short introduction to
  the lindblad master equation},\ }\href {https://doi.org/10.1063/1.5115323}
  {\bibfield  {journal} {\bibinfo  {journal} {AIP Advances}\ }\textbf {\bibinfo
  {volume} {10}},\ \bibinfo {pages} {025106} (\bibinfo {year}
  {2020})}\BibitemShut {NoStop}%
\bibitem [{\citenamefont {Nakanishi}\ and\ \citenamefont
  {Sasamoto}(2022)}]{nakanishi_pt_2022}%
  \BibitemOpen
  \bibfield  {author} {\bibinfo {author} {\bibfnamefont {Y.}~\bibnamefont
  {Nakanishi}}\ and\ \bibinfo {author} {\bibfnamefont {T.}~\bibnamefont
  {Sasamoto}},\ }\bibfield  {title} {\bibinfo {title} {{PT} phase transition in
  open quantum systems with {Lindblad} dynamics},\ }\href
  {https://doi.org/10.1103/PhysRevA.105.022219} {\bibfield  {journal} {\bibinfo
   {journal} {Phys. Rev. A}\ }\textbf {\bibinfo {volume} {105}},\ \bibinfo
  {pages} {022219} (\bibinfo {year} {2022})}\BibitemShut {NoStop}%
\bibitem [{\citenamefont {Shammah}\ \emph {et~al.}(2018)\citenamefont
  {Shammah}, \citenamefont {Ahmed}, \citenamefont {Lambert}, \citenamefont
  {De~Liberato},\ and\ \citenamefont {Nori}}]{PhysRevA.98.063815}%
  \BibitemOpen
  \bibfield  {author} {\bibinfo {author} {\bibfnamefont {N.}~\bibnamefont
  {Shammah}}, \bibinfo {author} {\bibfnamefont {S.}~\bibnamefont {Ahmed}},
  \bibinfo {author} {\bibfnamefont {N.}~\bibnamefont {Lambert}}, \bibinfo
  {author} {\bibfnamefont {S.}~\bibnamefont {De~Liberato}},\ and\ \bibinfo
  {author} {\bibfnamefont {F.}~\bibnamefont {Nori}},\ }\bibfield  {title}
  {\bibinfo {title} {Open quantum systems with local and collective incoherent
  processes: Efficient numerical simulations using permutational invariance},\
  }\href {https://doi.org/10.1103/PhysRevA.98.063815} {\bibfield  {journal}
  {\bibinfo  {journal} {Phys. Rev. A}\ }\textbf {\bibinfo {volume} {98}},\
  \bibinfo {pages} {063815} (\bibinfo {year} {2018})}\BibitemShut {NoStop}%
\end{thebibliography}
\end{document}